\documentclass{IEEEtran}

\usepackage{amsmath,bm,xfrac,mathtools}
\usepackage[scr=rsfs]{mathalfa}
\usepackage{hhline}
\usepackage{amssymb}  
\usepackage{graphicx, epsfig, epstopdf}
\usepackage{verbatim}
\usepackage{commath}
\usepackage{color}
\allowdisplaybreaks
\usepackage{theorem}
\usepackage{ifthen}
\usepackage{tcolorbox}
\usepackage{paralist}

\newcommand{\eps}{\varepsilon}
\renewcommand{\epsilon}{\varepsilon}

\newcommand{\foral}{\forall\,}
\newcommand{\timeder}[1]{\dot{#1}}

\definecolor{uiuc_orange}{RGB}{232,74,39}
\definecolor{uiuc_blue}{RGB}{17,40,74}

\theoremstyle{plain}

\newtheorem{Lemma}{Lemma}
\newtheorem{Proposition}{Proposition}

{\theorembodyfont{\rmfamily} \newtheorem{Remark}{Remark}}
{\theorembodyfont{\rmfamily} }
{\theorembodyfont{\rmfamily} }
{\theorembodyfont{\rmfamily} }

\newboolean{comments}

\setboolean{comments}{true}

\newcommand{\dl}[1]{\ifthenelse{\boolean{comments}}{\textcolor{red}{DL: #1}}{}}
\newcommand{\oa}[1]{\ifthenelse{\boolean{comments}}{\textcolor{blue}{ #1}}{}}
\newcommand{\proof}{\noindent{\it Proof: }}

\begin{document}
	
		\title{ Robust synchronization of electric power generators}
		
		\author{Olaoluwapo~Ajala,~\IEEEmembership{Student Member,~IEEE,}
			Alejandro~Dom\'{i}nguez-Garc\'{i}a,~\IEEEmembership{Member,~IEEE,}
			and~Daniel~Liberzon,~\IEEEmembership{Fellow,~IEEE}
			\thanks{The authors are with the Department of Electrical and Computer Engineering, University of Illinois at Urbana-Champaign, Urbana, IL 61801 USA. E-mail:\{ooajala2, aledan, psauer, liberzon\}@ILLINOIS.EDU.}
		\thanks{The work done by Olaoluwapo Ajala and Alejandro D. Dom\'{i}nguez-Garc\'{i}a was supported by the Advanced Research Projects Agency-Energy (ARPA-E), U.S. Department of Energy, within the NODES program, under Award DE-AR0000695. Daniel Liberzon's work was supported by the NSF grant CMMI-1662708 and the AFOSR grant FA9550-17-1-0236. A preliminary version of this paper will be presented at  the 57th IEEE Conference on Decision and Control (CDC 2018)}}
		
%
%
		
		

		\maketitle
		\begin{abstract}
			We consider the problem of synchronizing two electric power generators, one of which (the leader) is serving a time-varying electrical load, so that they can ultimately be connected to form a single power system. Each generator is described by a second-order reduced state-space model. We assume that the generator not serving an external load initially (the follower) has access to measurements of the leader's phase angle, corrupted by some additive disturbances. By using these measurements, and leveraging results on reduced-order observers with ISS-type robustness, we propose a procedure that drives (i) the angular velocity of the follower close enough to that of the leader, and (ii) the phase angle of the follower close enough to that of the point at which both systems will be electrically connected. An explicit bound on the synchronization error in terms of the measurement disturbance and the variations in the electrical load served by the leader is computed. We illustrate the procedure via numerical simulations.
		\end{abstract}
			

	\section{Introduction}
	Research into {synchronization} of dynamical systems originates in the 17th century study of pendulum clocks by Huygens and continues vigorously to this day, driven by theoretical interest and applications in mechanical and electrical systems, multi-agent coordination, teleoperation, haptics, and other fields.
	In the physics literature, the famous Pecora-Carroll synchronization scheme from~\cite{pecora-carroll} has generated a lot of activity, some of which was recently surveyed in~\cite{pecora-carroll-survey-2015}.
	In modern control-theoretic literature, tools that have been prominent in addressing synchronization problems are dissipativity theory~\cite{arcak-passivity-synchronization,AndrievskiiFradkov06,chopra-teleoperation} and observer design~\cite{NM97,pogromsky-nijmeijer-1998,lorenz}. In the context of electric power systems, Kuramoto-type models of coupled phase oscillators, which have been utilized in numerous areas since first proposed in \cite{Ku1975}, are also starting to be adopted to describe the behavior of inertia-less microgrids (see, e.g., \cite{dorfler-bullo-survey,FD-JWSP-FB:14a,JWSP-FD-FB:12u,ZoDo:16} and the references therein).	
	
	It is important to distinguish between two basic synchronization scenarios. The first one is when there is bidirectional exchange of information between systems that are already coupled (usually by mechanical or electrical forces) and are trying to achieve a common objective; see, e.g., \cite{dorfler-bullo-survey,WEITENBERG2018_1,Polyak2017}. The second scenario is when the flow of information is unidirectional: from a ``leader" to a ``follower."
	In this case, the follower and the leader are not physically coupled at first, but the follower is trying to emulate the behavior of the leader so as to attempt physical coupling. This second setting naturally arises in the problem of connecting an electrical generator to an electrical network, and it is the focus of this paper.
	
	In several applications, such as the one considered in this paper, it is important to guarantee an acceptable level of synchronization in the presence of errors affecting the measurements exchanged between the systems trying to synchronize. Such \emph{robust synchronization} problems have recently been receiving attention in the literature.
	Systems in Lurie form satisfying a passifiability assumption on the linear part were treated in~\cite{AndrievskiiFradkov06,FradkovAndrievsky_PRE08,FradkovAndrAnan_Aut15}.  The work reported in~\cite{pogromsky-nijmeijer-1998} establishes robustness of synchronization to uncertainties satisfying inequality constraints and relies on Lyapunov-based observer design. On the other hand, as discussed in~\cite{carroll-noise-05}, most known synchronization schemes  are quite sensitive to even small random noise, and very few general results addressing their robustness to bounded disturbances are presently available. The recent work~\cite{lorenz} addresses this problem using an ISS observer approach developed earlier in~\cite{qDES-obs}, which also serves as a conceptual basis for the synchronization scheme to be presented here.
	
	In the power systems literature, synchronization methods are categorized as manual, assisted-manual, or
	automatic, with each approach having unique benefits and limitations~\cite{schaefer2016,thompson2012}. In manual synchronization
	methods, an operator visualizes {the voltage, frequency, and phase differences}
	of the connection points using a synchronizing panel, manually adjusts the system controls to establish
	synchronization, and manually initiates a connection when the systems are synchronized. In assisted-manual
	methods, a supervisory relay is added to the manual synchronization method and is
	tasked with ensuring that two power systems cannot be connected unless they are synchronized. In automatic
	synchronization methods, synchronization relays are used, and the entire process of synchronization and
	connection is automated. Although the manual, assisted-manual, and automatic synchronization methods
	are well established in the power system literature, their robustness is not rigorously addressed. {Disturbances in the measured voltage, frequency and phase of the connection points can potentially result in damage to electrical components and propagation of disturbances across the power system~\cite{thompson2012}. In light of this, robust synchronization methods are necessary for electric power generators.}
	
	In this paper, we consider two power systems that are not electrically connected, with the ultimate goal of interconnecting them to form a single system with all its generators being synchronized. Here, we focus on the case when the first system, referred to as the leader, is comprised of one generator and one load, both of which are connected to a bus with voltage support; and the second system, referred to as the follower system, is comprised of a single generator. The objective then is to synchronize both systems, i.e., make the generators rotate at the same angular velocity, and make the voltage magnitude and phase angle of the point at which they will be interconnected match. Once these two objectives are achieved, it is possible to electrically connect the follower system to the leader system without causing large currents to flow across both systems, or causing mechanical components to break (see, e.g., \cite{thompson2012}).
	
	By assuming the load in the leader system is not varying too rapidly, we first show that a standard integral control stabilizes the angular velocity of the generator in the leader system.  {Then, by assuming the follower system has access to only voltage magnitude and phase measurements (but not angular velocity measurements) of the leader system, we show that even if the phase measurements are corrupted, due to, e.g., noise or a malicious cyber attack, the generator in the follower system will be able to bring its angular velocity close enough to that of the generator in the leader system.} As for phase synchronization, our procedure cannot guarantee that the phase difference will converge to within some small value around zero; in fact, the opposite is generally true---the phase difference will grow unbounded over time. In turns out, however, that this is not a problem in practice, since one just needs to wait until the phase difference is a multiple of $2\pi$ to physically interconnect both systems.
	
	A preliminary study of the basic control design and synchronization methods presented in this work was first conducted in~\cite{ajala_cdc2018}. However, the presentation given in this paper is more complete and includes additional results and formulations. A derivation of the generator model used in our analysis is included as an Appendix---although a library of models containing the one considered here was presented in \cite{Library2d}, here we include the assumptions used to further reduce the model to a more tractable form. Also, in this paper we present small-signal analysis results to validate claims made in~\cite{ajala_cdc2018} about the effect of load perturbations on phase variations. More complete numerical and analytical results are developed here compared to~\cite{ajala_cdc2018}; and finally, analytical results for the more general case when the generator damping function is phase-dependent are presented.
	
	The remainder of the paper is organized as follows. In Section~\ref{s-system-description}, we explain the problem considered in this work, i.e. synchronizing two electric power generators. We present the mathematical models used and discuss the assumptions made in the problem formulation. In Section~\ref{s-ctrlsync}, we propose and design tools---a feedback control law and a synchronization method---for solving the synchronization problem. In Section~\ref{s-results}, numerical results are presented to validate our proposed control law and synchronization method. In Section~\ref{s-phase-depend-damp}, we show that our proposed synchronization method {is applicable to a more general class of problems}, i.e. when the damping coefficients in the mathematical models are phase-dependent, rather than constant. In Section~\ref{s-conclusion}, concluding remarks are discussed, and in the Appendix, a derivation of the mathematical models used is presented.

	\section{System description}\label{s-system-description}

\begin{figure}[htbp!]
		\centerline{
			\includegraphics[width=0.9\columnwidth]{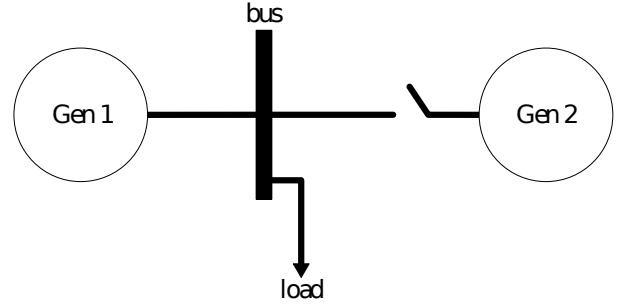}}
		\caption{Synchronization of two generators: a leader and a follower.
			\label{fig:1bus1gen}}
	\end{figure}

	We focus on the task of synchronizing two electric power generators, with the first one serving an electrical load via a node referred to as  the ``bus," and the second one trying to connect to the bus. The synchronization task is depicted in Fig. \ref{fig:1bus1gen}.
	
	Let $\omega_1$ denote the angular speed of the first generator (in electrical radians per second), let $\theta_1$ denote the absolute phase angle
	of generator~1,  and let $\delta_1$ denote its relative phase angle, both in radians. This means that
	\begin{align}\label{e-deltagen}
	\delta_1 := \theta_1-\omega_0t,
	\end{align}
	where $\omega_0$ denotes some nominal frequency; thus, we have $\dot{\theta}_1= \omega_1$, so that
	\begin{equation}\label{e-delta1-ode}
	\timeder{\delta}_1 = \omega_1 - \omega_0.
	\end{equation}
	The corresponding variables $\omega_2$, $\theta_2$, $\delta_2$ for the second generator are defined in the same way. The bus state variables are the voltage magnitude and the voltage angle for the bus. We denote by  $\theta_3$ the absolute phase angle of the bus voltage. We also define the relative phase angle of the bus voltage as
	\begin{align}\label{e-deltabus}
	\delta_3 := \theta_3-\omega_0t,
	\end{align}
	and we have $\dot{{\theta}}_3= \omega_3$, so that
	$
	\timeder{{\delta}}_3 = \omega_3 - \omega_0,
	$
	where $\omega_3$ is the frequency of the bus (in electrical radians per second).

	We consider the following second-order reduced model for the first generator, which is based on an assumption that the voltage support of the leader system maintains the voltage magnitude of the ``bus'' node at its rated value (see the Appendix for model derivation details):
	\begin{align}
	\timeder{\theta}_1 =&\ \omega_1,\label{eqn:powerangleD}\\
	\timeder{\omega}_1
	=&\ u_1 -\ell(t)- {D}_1^{(0)}\omega_1,
	\label{eqn:gen_ang_speed1}
	\end{align}
	where $u_1$ is the control input;
	\begin{equation}\label{e-load-new}
	\begin{split}
	\ell(t)=B_1(\theta_{13}(t)) + D_1(\theta_{13}(t))\cdot\dot{\theta}_{13}(t)
	\end{split}
	\end{equation}
	is the electrical load;
	\begin{equation}\label{e-Delta1-def}
	\begin{split}
	\theta_{13}(t):=\theta_1(t)-\theta_3(t)
	\end{split}
	\end{equation}
	is the difference between the absolute phase angles of the first generator and the bus;
	$B_1$ is a globally bounded and globally Lipschitz function given by
	\begin{equation}\label{e-B1-def}
	\begin{split}
	B_1(s):=K_1\sin(s)+X_1\sin(2s),
	\end{split}
	\end{equation}
	where $K_1$ is a positive constant and $X_1$ is a nonnegative constant \cite{Library2d};
	the damping function $D_1$ is a globally bounded and globally Lipschitz function {given by}
	\begin{equation}\label{e-D1-def}
	\begin{split}
	D_1(s)=C_1 \cos^2(s)+C_2\sin^2(s),
	\end{split}
	\end{equation}
	where $C_1$ and $C_2$ are nonnegative constants \cite{Library2d};
	and $D_1^{(0)}$ is a positive constant.

	From the generator dynamic model in \eqref{eqn:powerangleD}, \eqref{eqn:gen_ang_speed1}, and the definition~\eqref{e-Delta1-def} and the resulting relation
	\begin{equation}\label{e-dotDelta1}
	\begin{split}
	\dot\theta_{13}(t):&=\dot\theta_1(t)-\dot\theta_3(t)=\omega_1-\omega_3,
	\end{split}
	\end{equation}
	it is easy to see that the dynamical model for the bus takes the form
	\begin{align}
	\timeder\theta_3 =&\ \omega_3,\label{eqn:buspowerangle}\\
	\timeder\omega_3
	=&\ u_1 -\ell(t) - {D}_1^{(0)}\omega_1 - \ddot{\theta}_{13}(t) \label{eqn:genbus_ang_speed}.
	\end{align}

	The second-order reduced model for the second generator (before it is connected) is analogous to~\eqref{eqn:powerangleD}, \eqref{eqn:gen_ang_speed1} but with no electrical load term, i.e.,
	\begin{align}
	\timeder{\theta}_2 =&\ \omega_2,\label{eqn:buspowerangle_follow}\\
	\timeder{\omega}_2 =&\ u_2 - {D}^{(0)}_2{\omega_2}, \label{eqn:genbus_ang_speed_follow}
	\end{align}
	where $u_2$ is the control input and ${D}^{(0)}_2$ is a positive constant.
	
	The synchronization task consists in ensuring that the phase and angular speed of the second generator match those of the bus. Accordingly, from now on we refer to the bus modeled by~\eqref{eqn:buspowerangle}, \eqref{eqn:genbus_ang_speed} as the \emph{leader}, and the second generator modeled by~\eqref{eqn:buspowerangle_follow}, \eqref{eqn:genbus_ang_speed_follow} as the \emph{follower}.
	
	We assume that at the initial time $t_0$ (the time when our control strategy will be initialized), the first generator operates in steady state corresponding to some constant load $\bar\ell$. In view of the power balance equation~\eqref{e-load-new}, this means that $\theta_{13}(t_0)$ equals the solution $\bar\theta_{13}$ of the equation $\bar\ell=B_1(\bar\theta_{13})$, and that $\dot\theta_{13}(t_0)=\omega_1(t_0)-\omega_3(t_0)=0$. [Indeed, $\theta_{13}(t)\equiv\bar\theta_{13}$ is the unique solution of the ODE $\bar\ell=B_1(\theta_{13}(t))+ D_1(\theta_{13}(t))\cdot\dot{\theta}_{13}(t)$ starting at $\bar\theta_{13}$.]
	
	For $t\ge t_0$, we allow the load $\ell(t)$ to change, but assume that this change is constrained both in size and in speed, i.e., we assume that for some positive constants $\Delta_\ell$ and $\Delta_{\dot\ell}$ we have
	\begin{equation}\label{e-Delta-ell}
	\begin{split}
	|\ell(t)-\bar\ell|\le\Delta_\ell,\qquad |\dot\ell(t)|\le\Delta_{\dot\ell}.
	\end{split}
	\end{equation}
	{Letting $\Delta\theta\approx\theta_{13}(t)-\bar\theta_{13}$, $\Delta\dot\theta\approx\omega_1(t)-\omega_3(t)$, $\Delta\ell\approx\ell(t)-\bar\ell$, and $\Delta{\dot\ell}\approx\dot\ell(t)$ denote small perturbations about the initial values, one can show, by small-signal analysis, that
		\begin{equation}
		\begin{split}
		\frac d{dt}&\begin{pmatrix}\Delta\theta\\	 \Delta\dot\theta\end{pmatrix}=\begin{pmatrix}-\frac{1}{D_1(\bar{\theta}_{13})}\frac{\partial B_1(\bar{\theta}_{13})}{\partial{\theta}_{13}}\\
		\left(\frac{1}{D_1(\bar{\theta}_{13})}\frac{\partial B_1(\bar{\theta}_{13})}{\partial{\theta}_{13}}\right)^2\end{pmatrix}
		\Delta\theta\\&+\begin{pmatrix}
		-\frac{1}{D_1(\bar{\theta}_{13})}&0\\
		\frac{1}{\left(D_1(\bar{\theta}_{13})\right)^2}\frac{\partial B_1(\bar{\theta}_{13})}{\partial{\theta}_{13}}&-\frac{1}{D_1(\bar{\theta}_{13})}
		\end{pmatrix}\begin{pmatrix}\Delta\ell\\
		\Delta{\dot\ell}\end{pmatrix}
		\end{split}
		\end{equation}
		and if $\Delta_\ell$ and $\Delta_{\dot\ell}$ in~\eqref{e-Delta-ell} are sufficiently small then, at least on some finite time horizon, there exist positive constants $\Delta_\theta$ and $\Delta_{\dot\theta}$ such that
		\begin{equation}\label{e-Delta-theta}
		|\theta_{13}(t)-\bar\theta_{13}|\le\Delta_\theta,\; |\dot\theta_{13}(t)|=|\omega_1(t)-\omega_3(t)|\le\Delta_{\dot\theta}.
		\end{equation}
		We henceforth assume the existence of such constants $\Delta_\theta$ and $\Delta_{\dot\theta}$. }
	
	\subsubsection*{Signal measurements}
	
	We assume that a phasor-measurement unit (PMU) is used to measure the absolute angle, $\theta_{3}(t)$, of the ``bus" node, which is corrupted by a measurement disturbance, $d(t)$.\footnote{The voltage magnitude of the bus node is a known constant. As a result, the PMU does not need to measure it.} One major potential source of such a disturbance is \textit{spoofing} \cite{spoofing2013}, but it can also be due to a combination of several sources. Thus, phase measurements available to the follower take the form
	\begin{align}
	\theta_{3}(t) + d(t),
	\end{align} where $d(t)$ is an unknown disturbance, with $\theta_{3}(t) + d(t)\in[0,2\pi)$.\footnote{Note that if the unknown disturbance is caused by a spoofing attack on the GPS signal of the PMU, it might be possible to refine the upper bound on  $d(t)$. For example, in~\cite{spoofing2013}, it was shown that a spoofing attack can be engineered so as to perturb the  phase measurement provided by the PMU by as much as $0.25\pi$~rad without being detected; thus, in such a case, one could assume $d(t)\in(-0.25\pi,0.25\pi)$.}
	We also assume that the steady-state value $\bar\theta_{13}$ is known to the follower (through the knowledge of $\bar\ell$.)
	On the other hand, angular speed measurements are not available to the follower.
	
	
	Our goal is to achieve robust synchronization in the face of the unknown disturbance $d$, and to quantitatively characterize how the synchronization error is affected by the size of this disturbance.

	\section{Controlled synchronization}\label{s-ctrlsync}

	In this section, a feedback control law is designed for the leader and a synchronization method is developed for the follower system.

	\subsection{Control design and analysis}\label{ss-control-design}

	\subsubsection*{First generator and bus (leader)}
	Note that the first generator and the bus share the same control input. The purpose of this control is to drive the bus frequency $\omega_3(t)$ to the nominal frequency value $\omega_0$. In view of the second bound in~\eqref{e-Delta-theta}, if $\Delta_{\dot\theta}$ is small then this goal can also be approximately achieved by driving the angular speed $\omega_1(t)$ of the first generator to $\omega_0$. This suggests
	the following control input:
	\begin{align}
	u_1(t) =&\ -k\delta_1(t)
	=-k(\theta_1(t)-\omega_0t), \qquad k>0.\label{e-Tm1-def}
	\end{align}
	Since the dynamics of $\delta_1(t)$ are given by~\eqref{e-delta1-ode},
	it is easy to recognize in~\eqref{e-Tm1-def} a standard integral control law for making $\omega_1(t)$ asymptotically track the constant reference $\omega_0$.
	Under the action of this control,
	the first generator reduced-order model~\eqref{eqn:powerangleD}, \eqref{eqn:gen_ang_speed1} becomes:
	\begin{align}
	\timeder{\theta}_1 =&\ \omega_1,\label{eqn:powerangle5}\\
	\timeder{\omega}_1 =& -k\theta_1+ k\omega_0t - \ell(t) - {D}_1^{(0)}\omega_1.     \label{e-dotomega1-prelim}
	\end{align}
	
	To validate the control law~\eqref{e-Tm1-def}, we want to show that the solutions of the closed-loop system given by~\eqref{e-delta1-ode}, \eqref{eqn:powerangle5} and~\eqref{e-dotomega1-prelim} are bounded and that $\omega_1(t)$ is regulated to $\omega_0$ in an appropriate sense. To this end, it is convenient to
	rewrite the $(\omega_1,\delta_1)$-dynamics as follows:
	\begin{align*}
	\begin{pmatrix}\dot\omega_1\\
	\dot \delta_1\end{pmatrix}
	&=\begin{pmatrix}-D_1^{(0)}&-k\\
	1 &0\end{pmatrix}
	\begin{pmatrix}\omega_1\\
	\delta_1\end{pmatrix}-\begin{pmatrix}\ell(t)\\
	\omega_0\end{pmatrix},
	\end{align*}
	which we can view as a linear time-invariant system driven by a time-varying perturbation that creates a time-varying equilibrium at
	\begin{equation}\label{e-equil-TV}
	\begin{split}
	\omega_1=\omega_0,\qquad \delta_1=-\dfrac {\ell(t)+D_1^{(0)}\omega_0}k=:\delta_0(t)
	\end{split}
	\end{equation}
	(meaning that for each frozen time $t$, this is the equilibrium of the corresponding fixed affine system).
	Let us shift the center of coordinates to this time-varying equilibrium by defining
	\begin{equation}\label{e-shift-center}
	\bar\omega_1(t):=\omega_1(t)-\omega_0,\qquad  \bar\delta_1(t):=\delta_1(t)-\delta_0(t).
	\end{equation}
	Note that small values of $\bar\omega_1(t)$ correspond to $\omega_1(t)$ being regulated close to the nominal frequency $\omega_0$. The following result formally describes in what sense our controller achieves this goal.
	
	\begin{Proposition} For each $k>0$ there exist constants $c,\lambda>0$ such that
		the closed-loop system variables $\bar\omega_1$ and $\bar\delta_1$ satisfy the steady-state bound
		\begin{equation}\label{e-regulation-gain}
		\limsup_{t\to\infty}\left|\binom{\bar\omega_1(t)}
		{\bar \delta_1(t)}\right|\le \frac{c\Delta_{\dot\ell}}{\lambda k},
		\end{equation}
		where $\Delta_{\dot\ell}$ comes from~\eqref{e-Delta-ell}.
	\end{Proposition}

	\proof	In the new coordinates $(\bar\omega_1,\bar\delta_1)$, the closed-loop dynamics becomes
	\begin{align}
	\begin{pmatrix}\dot{\bar\omega}_1\\
	\dot{\bar\delta}_1\end{pmatrix}&=\begin{pmatrix}\dot\omega_1\\
	\dot\delta_1\end{pmatrix}-\begin{pmatrix}0\\
	\dot{\delta}_0(t)\end{pmatrix}\nonumber\\
	&=
	\begin{pmatrix}-D_1^{(0)}&-k\\
	1 &0\end{pmatrix}
	\begin{pmatrix}\bar\omega_1\\
	\bar\delta_1\end{pmatrix}+\begin{pmatrix}0\\
	\nu(t)\end{pmatrix},\label{e-affine-TI}
	\end{align}
	where
	\begin{equation}\label{e-nu}
	\begin{split}
	\nu(t):=\dfrac {\dot\ell(t)}k.
	\end{split}
	\end{equation}
	
	Since the matrix
	\begin{equation}\label{e-A}
	A:=\begin{pmatrix}-D_1^{(0)}&-k\\
	1 &0\end{pmatrix}
	\end{equation}
	is Hurwitz for every $k>0$, it is clear that closed-loop solutions are bounded and converge to a  neighborhood of the time-varying equilibrium~\eqref{e-equil-TV}; the size of this neighborhood is determined by the size of the perturbation $\nu(t)$. To make this more precise, note that since $A$ is Hurwitz, there exist constants $c,\lambda>0$ (which depend on $k$) such that for all $t$ we have\footnote{Here $\|\cdot\|$ stands for the induced matrix
		norm corresponding to the Euclidean norm.}
	\begin{equation}\label{e-commar-TI}
	\big\|e^{A t }\big\|\le ce^{-\lambda t }.
	\end{equation}
	Computation of $c$ and $\lambda$ is addressed in Section~\ref{ss-param-eval}.
	Our system~\eqref{e-affine-TI} is the LTI system $\dot x=Ax$ driven by the perturbation~\eqref{e-nu} which,  in view of the second bound in~\eqref{e-Delta-ell},
	satisfies  $$|\nu(t)|\le\frac{\Delta_{\dot\ell}}k\qquad \foral t\ge 0.$$
	It is well known and straightforward to derive
	that ${c}/{\lambda}$ is an upper bound on the system's $\mathcal L_\infty$-induced gain, and that the following bound holds for all solutions:
	$$
	\left|\binom{\bar\omega_1(t)}
	{\bar \delta_1(t)}\right|\le c e^{-\lambda t}\left|\binom{\bar\omega_1(0)}
	{\bar \delta_1(0)}\right|+\frac{c}{\lambda}\frac{\Delta_{\dot\ell}}k
	\qquad \foral t\ge 0.
	$$
	In particular, $c{\Delta_{\dot\ell}}/({\lambda}k)$ is the ultimate bound on the norm of the solution in steady state, as claimed in~\eqref{e-regulation-gain}.~\IEEEQEDhere

	\subsubsection*{Second generator (follower)}
	
	For the follower (second generator) described by~\eqref{eqn:buspowerangle_follow}, \eqref{eqn:genbus_ang_speed_follow}, we would like to define the control input $u_2(t)$ so as to make the angular speed $\omega_2(t)$ synchronize with the bus frequency $\omega_3(t)$. Since in view of the second bound in~\eqref{e-Delta-theta} the frequencies $\omega_3(t)$ and $\omega_1(t)$ are close to each other, it is reasonable to base the design of $u_2$ on the (somewhat simpler) dynamics of the first generator instead of those of the bus. Let us use~\eqref{e-load-new} to rewrite the equation~\eqref{e-dotomega1-prelim}
	as
	\begin{equation}\label{eqn:gen_ang_speed5}
	\begin{split}
	\timeder{\omega}_1=& -k\theta_1 + k\omega_0t - B_1({\theta}_{13}(t))\\&- D_1(\theta_{13}(t))\cdot\dot\theta_{13}- {D}_1^{(0)}\omega_1.
	\end{split}
	\end{equation}
	We can make the dynamics~\eqref{eqn:genbus_ang_speed_follow} of $\omega_2$ approximately match these dynamics of $\omega_1$ by doing the following:
	(i)~approximating $\theta_1(t)$ (which is not available to the follower) by $\theta_3(t)+d(t)+\bar\theta_{13}$---this makes sense since $\theta_3(t)+d(t)$ are the approximate measurements of $\theta_3(t)$ available to the follower, and $\bar\theta_{13}$ approximates the difference $\theta_{13}(t)=\theta_1(t)-\theta_3(t)$ in the sense of the first bound in~\eqref{e-Delta-theta} and is also available to the follower; (ii)~approximating $B_1({\theta}_{13}(t))$ by
	$B_1({\bar\theta}_{13})$; (iii)~correcting the difference between the damping constants $D_1^{(0)}$ and $D_2^{(0)}$; and (iv)~ignoring the term $D_1(\theta_{13}(t))\cdot\dot\theta_{13}$ which is bounded by virtue of~\eqref{e-D1-def} and~\eqref{e-Delta-theta}.
	This suggests the following control input:
	\begin{align*}
	u_2(t) =&\  - k\big(\theta_3(t)+d(t)+\bar\theta_{13}\big)+ k\omega_0t\ - B_1({\bar\theta}_{13})\\&+\big({D}^{(0)}_2 - {D}_1^{(0)}\big){\omega_2(t)}.
	\end{align*}
	We can then write the closed-loop dynamics of the follower as
	\begin{align}
	\timeder{\theta}_2 =&\ \omega_2,\label{eqn:buspowerangle_follow1}\\
	\timeder{\omega}_2 =& - k\big(\theta_3(t)+d(t)+\bar\theta_{13}\big)+ k\omega_0t - B_1({\bar\theta}_{13})- {D}^{(0)}_1{\omega_2}. \label{eqn:genbus_ang_speed_follow1}
	\end{align}
	This choice of control for the follower will be validated by the synchronization analysis given next.
	
	\begin{Remark}
		The above control design for the follower is not dependent on the particular form of the control $u_1$ for the leader, but only on the fact that this control depends just on the angle $\theta_1$ and not on the angular velocity $\omega_1$, so that the follower can approximately reconstruct this control (modulo the disturbance). We also see that the exact nature of the damping term in the follower model is not important because it is canceled by control.
	\end{Remark}
	
	\subsection{Synchronization analysis}\label{ss-sync}
	
	Since we are interested in synchronizing the angular velocity $\omega_2$ of the follower to the frequency $\omega_3$ of the leader, we consider the synchronization error
	\begin{equation}\label{e-error-def}
	e(t)\coloneqq\omega_2(t)-\omega_3(t).
	\end{equation}
	The following result characterizes the quality of synchronization in terms of the size of the disturbance $d(t)$, the control gain $k$, the damping coefficient ${D}_1^{(0)}$, and the various constants appearing in~\eqref{e-B1-def}, \eqref{e-D1-def}, and~\eqref{e-Delta-theta}.

	\begin{Proposition}\label{p-sync}
		Along the closed-loop dynamics of the leader and the follower defined in Section~\ref{ss-control-design}, the synchronization error~\eqref{e-error-def} satisfies the steady-state bound
		{\begin{equation}\label{e-gain-combined}
			\begin{split}
			\limsup_{t\to\infty}|{e}(t)|\le &\Big(k\limsup_{t\to\infty}| d(t)|\\&\,+(C_1+C_2+{D}_1^{(0)})\Delta_{\dot\theta}\\&\ +(k+K_1+2X_1)\Delta_\theta
			\Big)\dfrac1{{D}_1^{(0)}}.
			\end{split}
			\end{equation}}
	\end{Proposition}
	
	This bound shows, in particular, that the gain from the measurement disturbance $d$ to the synchronization error $e$  is proportional to the control gain $k$, thus decreasing $k$ reduces the effect of this disturbance on synchronization. On the other hand, decreasing $k$ has a negative effect on closed-loop stability of the first generator, as can be seen from the eigenvalues of the matrix $A$ defined in~\eqref{e-A} and from the bound~\eqref{e-regulation-gain}.  {This suggests that, to mitigate the effect of this disturbance, we may want to (temporarily) reduce the control gain $k$ during the synchronization stage.

		\proof	We find it convenient to split $e$ as
		\begin{equation}\label{e-error-split}
		e=(\omega_2-\omega_1)+(\omega_1-\omega_3)=:e_{21}+e_{13}
		\end{equation}
		and analyze the two components separately. For $e_{13}$, we already have the second bound from~\eqref{e-Delta-theta} which says that
		\begin{equation}\label{e-e13}
		|e_{13}(t)|\le\Delta_{\dot\theta}.
		\end{equation}
		For $e_{21}$, using~\eqref{eqn:genbus_ang_speed_follow1}, \eqref{eqn:gen_ang_speed5}, and~\eqref{e-Delta1-def} we have (suppressing all time arguments for simplicity)
		\begin{equation}\label{e-tilde-e21}
		\begin{split}
		\dot{e}_{21}=&\ \dot{\omega}_2-\dot{\omega}_1\\
		=&\
		B_1({\theta}_{13})-
		B_1({\bar\theta}_{13})+D_1(\theta_{13})\cdot\dot\theta_{13}- {D}_1^{(0)}{e}_{21}+k(\theta_{13}-\bar\theta_{13})-kd.
		\end{split}
		\end{equation}
		Let us define the candidate Lyapunov function
		$$
		V({e}_{21}):=\frac 12 {e}_{21}^2.
		$$
		Its derivative along solutions of~\eqref{e-tilde-e21} satisfies the inequality
		\begin{equation}\label{e-V-dot}
		\begin{split}
		\dot V \le&
		-{D}_1^{(0)}{e}_{21}^2+\Big(k|\theta_{13}-\bar\theta_{13}|+k|d|+|B_1({\theta}_{13})\\&-
		B_1({\bar\theta}_{13})|+|D_1(\theta_{13})|\cdot|\dot\theta_{13}|\Big)|e_{21}|.
		\end{split}
		\end{equation}
		Recall that ${D}_1^{(0)}>0$. By the first bound in~\eqref{e-Delta-theta} we have $|\theta_{13}-\bar\theta_{13}|\le\Delta_\theta$. Furthermore, since $B_1$ defined in~\eqref{e-B1-def} is globally Lipschitz with Lipschitz constant $K_1+2X_1$, we also have $|B_1({\theta}_{13})-
		B_1({\bar\theta}_{13})|\le (K_1+2X_1)\Delta_\theta$. Finally, $D_1$ defined in~\eqref{e-D1-def} is globally bounded by $C_1+C_2$ which, combined with the second bound in~\eqref{e-Delta-theta}, gives $|D_1(\theta_{13})|\cdot|\dot\theta_{13}|\le (C_1+C_2)\Delta_{\dot\theta}$. Plugging all these bounds into~\eqref{e-V-dot}, we obtain
		\begin{align*}
		\dot V\le&
		-{D}_1^{(0)}{e}_{21}^2+\Big(k|d|+(k+K_1+2X_1)\Delta_\theta\\&+
		(C_1+C_2)\Delta_{\dot\theta}\Big)|e_{21}|\\=&
		-{D}_1^{(0)}|{e}_{21}|\Big(|e_{21}|-\\
		&-\frac{k|d|+(k+K_1+2X_1)\Delta_\theta+
			(C_1+C_2)\Delta_{\dot\theta}}{{D}_1^{(0)}}
		\Big),
		\end{align*}
		which yields
		\begin{align*}
		|{e}_{21}|&>\frac{k|d|+(k+K_1+2X_1)\Delta_\theta+
			(C_1+C_2)\Delta_{\dot\theta}}{{D}_1^{(0)}}\\&
		\Rightarrow\quad \dot V<0.
		\end{align*}
		The standard ISS analysis (see, e.g., \cite{sontag-coprime}) now implies that
		$e_{21}(t)$ stays bounded and satisfies the ultimate bound
		\begin{align*}
		\limsup_{t\to\infty}|{e}_{21}(t)|\le &\Big(k\limsup_{t\to\infty}| d(t)|+
		(C_1+C_2)\Delta_{\dot\theta}\\&+(k+K_1+2X_1)\Delta_\theta\Big)\dfrac1{{D}_1^{(0)}}.
		\end{align*}
		Combining this with~\eqref{e-error-split} and~\eqref{e-e13}, we arrive at the desired bound~\eqref{e-gain-combined}.~\IEEEQEDhere 
		
		Proposition~\ref{p-sync} can be viewed as a special case of the results in~\cite{qDES-obs} on reduced-order observers with ISS-type robustness.

		\subsubsection*{Synchronization procedure}
		
		In addition to angular velocity synchronization, phase synchronization is also important. The phase $\theta_2$ will evolve according to~\eqref{eqn:buspowerangle_follow1}, which comes from the physics of the system but was not explicitly taken into account in the above procedure. Due to the imperfect frequency synchronization caused by the disturbance, the phase difference $\theta_2-\theta_3$ will ``drift" and there will be a time when $\theta_2(t)-\left(\theta_3(t)+d(t)\right)$ will become close to an integer multiple of $2\pi$. The idea is that we will detect when this happens at the follower's side by looking at the measurements $\theta_3+d$ and comparing them with $\theta_2$, and at that moment we will connect the second generator.\footnote{For some disturbances that oscillate around 0, it is possible in principle that $\theta_2(t)-\left(\theta_3(t)+d(t)\right)$ will remain bounded and will never become a multiple of $2\pi$. However, for most disturbances---including constant-sign offsets arising from spoofing~\cite{spoofing2013}---the procedure is guaranteed to work.}
		
		The previous synchronization analysis can also be used to upper-bound the time that one must wait before satisfactory angular velocity matching is achieved. Indeed, the calculations given in the proof of Proposition~\ref{p-sync} imply that, for an arbitrary choice of $\eps>0$, we have
		\begin{align}
		&|e_{21}|\ge\
		\frac{(C_1+C_2)\Delta_{\dot\theta}}{{D}_1^{(0)}}(1+\eps)\nonumber\\&+\frac{k\sup_{0\le s\le t}|d(s)|+(k+K_1+2X_1)\Delta_\theta}{{D}_1^{(0)}}(1+\eps)
		\label{e-ball-inflated}\\&
		\Rightarrow\quad \dot V\le -{D}_1^{(0)}\frac{2\eps}{1+\eps}V.\label{e-Vdot-exp}
		\end{align}
		Therefore, as long as the inequality~\eqref{e-ball-inflated} is satisfied, the bound~\eqref{e-Vdot-exp} implies that $e_{21}(t)$ decreases exponentially according to
		$$
		|e_{21}(t)|\le e^{-{D}_1^{(0)}\textstyle{\frac{\eps}{1+\eps}}t}|e_{21}(0)|=
		e^{-{D}_1^{(0)}\textstyle{\frac{\eps}{1+\eps}}t}\omega_0
		$$
		where the second equality follows by assuming that, at time 0, the first generator is operating in steady state so that its angular velocity $\omega_1$ is close to the nominal value $\omega_0$, while the second generator is at rest so that $\omega_2(0)=0$, and recalling that $e_{21}=\omega_2-\omega_1$. Combined with~\eqref{e-error-split} and~\eqref{e-e13}, this gives us a (possibly quite conservative) estimate on the time before the mismatch between the angular velocities of the leader and the follower becomes close to its steady-state value.

		\subsubsection*{Post-synchronization system}As the leader and follower are synchronized and connected to form a single power system, the models governing the behavior of the two generators change. The dynamics of the first generator are now described by
		\begin{equation}
		\begin{split}
		\timeder{\theta}_1 =&\ \omega_1,\\
		\timeder{\omega}_1
		=&\ u_1 -B_1(\theta_{13}(t)) - D_1(\theta_{13}(t))\cdot\dot{\theta}_{13}(t) - {D}_1^{(0)}\omega_1,
		\end{split}
		\label{eqn:gen_1_sync}
		\end{equation} the dynamics of the second generator are described by
		\begin{equation}
		\begin{split}
		\timeder{\theta}_2 =&\ \omega_2,\\
		\timeder{\omega}_2
		=&\ u_2 -B_2(\theta_{23}(t)) - D_2(\theta_{23}(t))\cdot\dot{\theta}_{23}(t)- {D}_2^{(0)}\omega_2,
		\end{split}
		\label{eqn:gen_2_sync}
		\end{equation} and the power balance equation for the system is
		\begin{equation}\label{e-load-newer}
		\begin{split}
		\ell(t)=&\ B_1(\theta_{13}(t)) + D_1(\theta_{13}(t))\cdot\dot{\theta}_{13}(t)\\ & + B_2(\theta_{23}(t))+ D_2(\theta_{23}(t))\cdot\dot{\theta}_{23}(t),
		\end{split}
		\end{equation}	
		where $B_2$ and ${D}_2$ are globally bounded and globally Lipschitz functions, taking the same form as $B_1$ and ${D}_1$, and
		\begin{equation}\label{e-Delta1-def2}
		\begin{split}
		\theta_{23}(t):=\theta_2(t)-\theta_3(t)
		\end{split}
		\end{equation}
		is the difference between the absolute phase angles of the second generator and the bus.
		
		Let $\alpha_1,\alpha_2\in[0,1]$ denote participation factors of the leader and follower, respectively, where $\alpha_1+\alpha_2=1$, and let $z\in\mathbb{R}$ denote the automatic generation control variable (see \cite{wood1984power}, pp. 345--356, for more details). If the leader and follower are successfully synchronized, interconnected, and the system states approach a stable equilibrium, the control input of the leader and follower can be modified to ensure that the power consumed by the electrical load is shared according to participation factors $\alpha_1$ and $\alpha_2$, respectively, using the following control equations \cite{sauer2006power}:
		\begin{equation}
		\begin{split}
		\dot{z} =& -\left(\frac{{D}_1^{(0)}\omega_1(t)+{D}_2^{(0)}\omega_2(t)}{{D}_1^{(0)}+{D}_2^{(0)}}-\omega_0\right),\\
		u_1(t) =&\ \alpha_1 z(t),\quad
		u_2(t) =\ \alpha_2 z(t),\label{e-ctrl-input}
		\end{split}
		\end{equation}
		{These post-synchronization system dynamics and control will be used for generating the numerical results in the next section.}
		
		\section{Numerical results}
		\label{s-results}
		In this section, parameters for the proposed control law and synchronization method are evaluated, and numerical validations of both techniques are presented. The numerical results are developed as follows: with initial conditions of the leader system set to an equilibrium state and that of the follower system set to zero, the simulation starts at time $t = 0$~s with the electrical load at a nominal value of $0.5$ pu, where ``pu'' denotes per-unit.\footnote{System quantities expressed in per-unit have been normalized as fractions of a defined base quantity, and the rated value of the system quantity is usually chosen as the base quantity. In other words, for a system whose rated power capacity and voltage are 10 W and 480 V, respectively, a power measurement of 0.5 pu is equivalent to 5 W, and a voltage measurement of 1 pu is equivalent to 480 V \cite{Kundur1994}.} At time $t = 5$ s, the load is perturbed about the nominal value, with the change in size and speed constrained to $|\ell(t)-0.5|\le\Delta_\ell$ and $|\dot\ell(t)|=\Delta_{\dot\ell}$, respectively, where $\Delta_{\ell}$ and $\Delta_{\dot\ell}$ are positive constants. Using a base power of $2.2$~MW for the system, a base voltage amplitude of $480$~V for the generators, and a base voltage amplitude of $230$~kV for the bus, the model parameters are: $k=0.01$, $\omega_0=120\pi$ rad/s, $D_1^{(0)}=D_2^{(0)}=0.0531$ s/rad, $\bar{\ell}=0.5$~pu, $K_1=0.6434$~pu, $K_2=0.4167$~pu, $X_1=0.0742$~pu, $X_2=0.0742$~pu, $C_1=0.0656$~pu, $C_2=0.00548$~pu, and $\bar\theta_{13}=0.7245$~rad.

		\subsection{Parameter evaluation}\label{ss-param-eval}
		The values of $\lambda$ and $c$ in~\eqref{e-commar-TI} can be easily estimated as follows.
		The eigenvalues of $A$ are
		$$
		\lambda_{1,2}(A)=\frac{-D_1^{(0)}\pm\sqrt{\big(D_1^{(0)}\big)^2-4k}}2.
		$$
		To simplify calculations, let us assume that the control gain is chosen to satisfy
		$
		k\ge \big(D_1^{(0)}\big)^2/4
		$
		so that the eigenvalues of $A$ are complex with real parts
		$
		-\frac12{D_1^{(0)}}
		$.
		Then, we can take the stability margin (i.e., exponential decay rate) $\lambda$ appearing in~\eqref{e-commar-TI} to be
		$$
		\lambda:=\frac12 D_1^{(0)}.
		$$
		(Note that for values of $k$ closer to 0 the stability margin would decrease.)
		To calculate the overshoot constant $c$ in~\eqref{e-commar-TI}, we can look for a matrix $P=P^T>0$ which satisfies the Lyapunov inequality
		\begin{equation}\label{e-Lyap-ineq-TI}
		PA+A^TP\le-2\lambda P.
		\end{equation}
		Then, $A$ has its overshoot constant $c$ upper-bounded by $\sqrt{\lambda_{\text{max}}(P)/\lambda_{\text{min}}(P)}$. It can be verified that one choice of $P$ satisfying~\eqref{e-Lyap-ineq-TI} is $$
		P=\begin{pmatrix}
		1 & \frac12{D_1^{(0)}}\\
		\frac12{D_1^{(0)}} & k
		\end{pmatrix}
		$$
		(this actually gives $PA+A^TP=-D_1^{(0)}P$).
		Its eigenvalues are
		$$
		\lambda_{1,2}(P)=\frac{k+1\pm\sqrt{(k-1)^2+\big(D^{(0)}_1\big)^2}}2.
		$$
		If we fix some value of control gain $k>\big(D^{(0)}_1\big)^2/4$ (strict inequality is needed to have $P>0$), we obtain the following estimate for $c$:
		$$
		c=\sqrt{\dfrac{k+1+\sqrt{(k-1)^2+\big(D^{(0)}_1\big)^2}}{k+1-
				\sqrt{(k-1)^2+\big(D^{(0)}_1\big)^2}}}.
		$$
		(To refine this result, we can search for a matrix $P$ that gives the smallest value of $c$.)  Utilizing the formulas derived above and the chosen model parameters, we have that $c=10.3796$ and $\lambda=0.0266$.
		
		\subsection{Control performance analysis}
		For the leader to be in compliance with the IEEE 1547 standard \cite{IEEE1547}, we must have that $\abs{\omega_3(t)-\omega_0}\leq\pi$, and this should be enforced throughout system operation. Accordingly, effects of various model parameters on control performance are analyzed numerically.
		
		Firstly, the relation between bounds in \eqref{e-Delta-ell} and \eqref{e-Delta-theta} is investigated. The numerical results depicted in Figs. \ref{fig:Delta_theta} and \ref{fig:Delta_dot_theta} suggest that there is a strong coupling between $\Delta_{\theta}$ and variables $\Delta_{\ell}$ and $\Delta_{\dot{\ell}}$, and between $\Delta_{\dot{\theta}}$ and $\Delta_{\ell}$. However, there is a weak coupling between $\Delta_{\dot{\theta}}$ and $\Delta_{\dot{\ell}}$.
		\begin{figure}[htbp!]
			\centering
			\includegraphics[width=0.8\columnwidth]{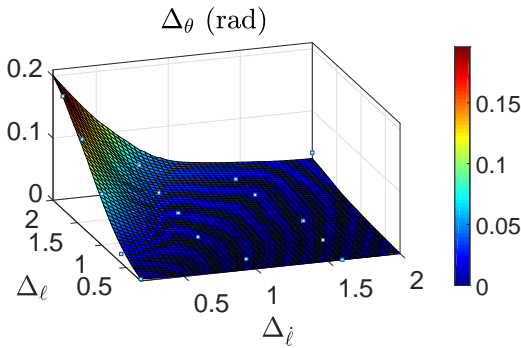}
			\caption{ Effect of $\Delta_{\ell}$ and $\Delta_{\dot{\ell}}$ on $\Delta_{\theta}$.
				\label{fig:Delta_theta}}
		\end{figure}
		\begin{figure}[htbp!]
			\centering
			\includegraphics[width=0.8\columnwidth]{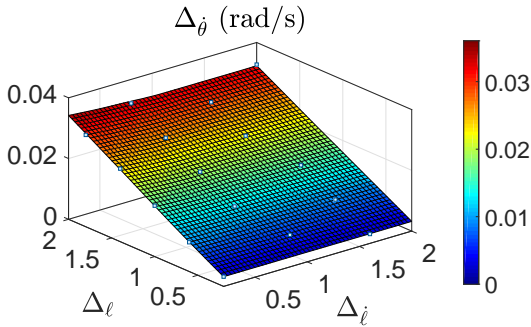}
			\caption{ Effect of $\Delta_{\ell}$ and $\Delta_{\dot{\ell}}$ on $\Delta_{\dot{\theta}}$.
				\label{fig:Delta_dot_theta}}
		\end{figure}
		Secondly, as depicted in Fig.~\ref{fig:freq_dev}, the deviation of the bus frequency from nominal value is investigated and compared to the bound required by the IEEE 1547 standard, i.e. $\abs{\omega_3(t)-\omega_0}\leq\pi$. The effects of $\Delta_{\ell}$ and $\Delta_{\dot{\ell}}$ on the frequency of the bus was also analyzed.
		\begin{figure}[htbp!]
			\centering
			\includegraphics[width=0.8\columnwidth]{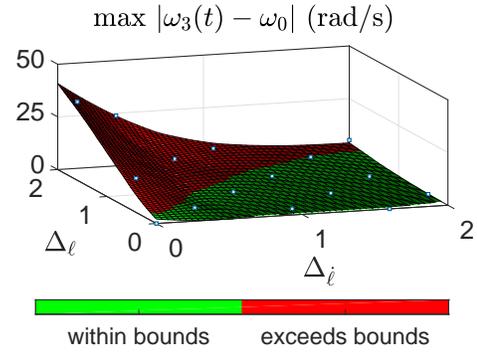}
			\caption{Frequency deviation relative to the bounds required by the IEEE 1547 standard ($k=0.01$).
				\label{fig:freq_dev}}
		\end{figure}
		{The results show that, for each fixed value of $\Delta_{\dot{\ell}}$, the controller performance improves when $\Delta_{\ell}$ decreases, and for each fixed value of $\Delta_{\ell}$, the controller performance improves when $\Delta_{\dot{\ell}}$ increases. Although Fig.~\ref{fig:freq_dev} suggests a weaker coupling between  $\Delta_{\dot{\ell}}$ and the controller performance, this result appears to contradict the bound in \eqref{e-regulation-gain}. However, it is important to note that this bound also takes into account the effects of phase deviations from a nominal value.}
		
		Finally, numerical results are presented in Fig. \ref{fig:norm} to validate the analytical result in \eqref{e-regulation-gain}. For all values of $k$ and $\Delta_{\dot{\ell}}$, we see that $\left|\begin{pmatrix}
		\bar{\omega}_1(t)\;\;\;\bar{\delta}_1(t)
		\end{pmatrix}^\top\right|\le\dfrac{c\Delta_{\dot{\ell}}}{\lambda k}$. Although the theoretical bound appears to be very conservative, the norm of the states is observed to be within these bounds, as expected.
		\begin{figure}[htbp!]
			\centering
			\includegraphics[width=0.8\columnwidth]{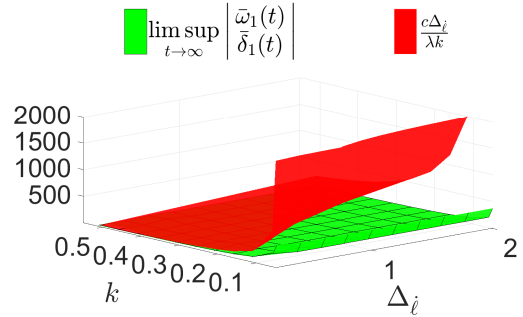}
			\caption{ Norm of states ($\Delta_{\ell}=0.1$).
				\label{fig:norm}}
		\end{figure}
		
		\subsection{Synchronization performance analysis}
		In order to further validate the claims of Proposition \ref{p-sync}, we investigated the effects of disturbance $d(t)$ and control gain $k$ on the synchronization error using numerical simulations. Taking into account the largest admissible synchronization error specified by IEEE standards listed in~\cite{thompson2012}, i.e. $|e(t)|\leq 0.134\pi$  rad/s, we also observed the performance of our proposed synchronization method.
		
		Utilizing results in \cite{spoofing2013} for the maximum phase angle error resulting from spoofing attacks, i.e. $d(t)\in(-0.25\pi,0.25\pi)$ rad, and simulating the effects of increasing $k$ on the synchronization error, we are able to observe that the synchronization error increases with control gain $k$ and disturbance $d(t)$, as shown in Fig. \ref{fig:sync_error_kd}. This is consistent with the analytical results in \eqref{e-gain-combined}. Comparing Figs. \ref{fig:norm} and \ref{fig:sync_error_kd}, it may be said that an increase in control gain $k$ improves controller performance, but at the cost of lowering robustness of the synchronization method. Also, Fig. \ref{fig:sync_error_kd} suggests that satisfactory performance of the proposed synchronization scheme is achieved when $k\in(0,0.4)$, and Fig. \ref{fig:freq_dev} suggests that satisfactory control performance is achieved when $k=0.01$ (an order of magnitude less than $0.4$). Collectively, these observations imply that, if other parameters are chosen consistently, the same value of $k$ can be used to achieve satisfactory performance of the control scheme and the synchronization scheme. In other words, there is no conflict between the requirement that k be large enough for satisfactory performance of the control scheme and the requirement that k be small enough for satisfactory performance of the proposed synchronization scheme.
		
		{In Figs. \ref{fig:sync_error}--\ref{fig:power_dev_follow}, the synchronization error, bus frequency and generator output power of the post-synchronization system are depicted for three constant disturbance values, $d(t)=0.125\pi$~rad, $d(t)=0.25\pi$~rad, and $d(t)=0.5\pi$~rad.\footnote{We also considered non-constant disturbances oscillating within the same magnitude limits, and observed even better results, suggesting that constant disturbances present a worst-case scenario.} The leader and follower are interconnected only when \begin{inparaenum}[(i)] \item the observed phase difference of the connection points, $\abs{\theta_2(t)-\theta_3(t)-d(t)}$, is a multiple of $2\pi$ and \item the synchronization error is within the admissible limits, i.e. $|e(t)|\leq 0.134\pi$  rad/s. \end{inparaenum} After the leader and follower were synchronized and interconnected, the post-synchronization control in \eqref{e-ctrl-input} was applied to the post-synchronization system at around $t=400$ s.  The values $\Delta_{\ell}=0.01$, $\Delta_{\dot{\ell}}=0.01$, and $k=0.01$ are used and the participation factors are $\alpha_1=\alpha_2=0.5$.}
		
		{Utilizing the main result of Proposition \ref{p-sync}, the steady state bounds for the synchronization error are given in Table \ref{tab:steady-state_bound}. Comparing these bounds to the admissible limits, i.e. $|e(t)|\leq 0.134\pi$  rad/s, we expect that the leader and follower will synchronize when $d(t)=0.125\pi$~rad.}
		\begin{center}
			\begin{table}[h!t!]
				\centering
				\caption{Theoretical Steady State Bounds for Synchronization Error ($e(t)$)}
				\begin{tabular}{|c |c|}
					\hline
					{$d(t)$} & {Theoretical Bounds for $e(t)$} \\ \hhline{|=|=|}
					$0.125\pi$ rad & $0.131\pi$ rad/s \\ \hline
					$0.25\pi$ rad & $0.1546\pi$ rad/s \\ \hline
					$0.5\pi$ rad & $0.2017\pi$ rad/s \\ \hline
				\end{tabular} \label{tab:steady-state_bound}
			\end{table}
		\end{center}
		
		Examining the results depicted in Fig. \ref{fig:sync_error}, we observed that for: \begin{inparaenum}[(i)]
			\item $d(t)=0.125\pi$~rad, the leader and follower successfully synchronized around $t = 200$ s,
			\item $d(t)=0.25\pi$~rad, the leader and follower successfully synchronized around $t = 260$ s, and
			\item  $d(t)=0.5\pi$~rad, the leader and follower fail to synchronize.
		\end{inparaenum} This is consistent with the synchronization error bounds listed in Table \ref{tab:steady-state_bound} in the sense that it predicts that the leader and follower will synchronize when $d(t)=0.125\pi$~rad. The results depicted in Fig. \ref{fig:sync_error} suggest that the proposed synchronization method is robust to large disturbances in phase measurements, even if the disturbance is as large as the maximum resulting from spoofing attacks. Also, as depicted in Figs.~\ref{fig:power_dev_lead} and \ref{fig:power_dev_follow}, we observed that after successful synchronization, the leader and follower generator share the load equally, according to the participation factors.
		
		\begin{figure}[htbp!]
			\centering
			\includegraphics[width=0.95\columnwidth]{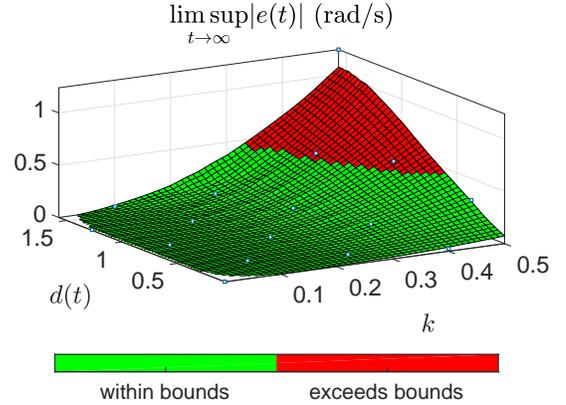}
			\caption{ The synchronization error limit, as described in \eqref{e-gain-combined}, relative to bounds prescribed in \cite{thompson2012}. $d(t)$ represents constant disturbance values.
				\label{fig:sync_error_kd}}
		\end{figure}
		\begin{figure}[htbp!]
			\centering
			\includegraphics[width=0.8\columnwidth]{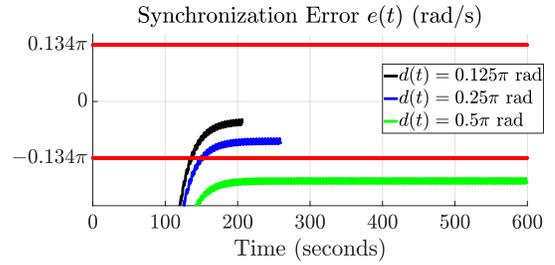}
			\caption{ Synchronization error relative to bounds provided in \cite{thompson2012}.
				\label{fig:sync_error}}
		\end{figure}
		\begin{figure}[htbp!]
			\centering
			\includegraphics[width=0.85\columnwidth]{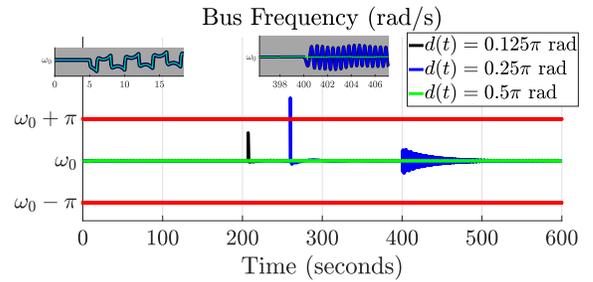}
			\caption{Bus frequency relative to bounds provided in \cite{IEEE1547}.\label{fig:bus_freq}}
		\end{figure}
		\begin{figure}[htbp!]
			\centering
			\includegraphics[width=0.75\columnwidth]{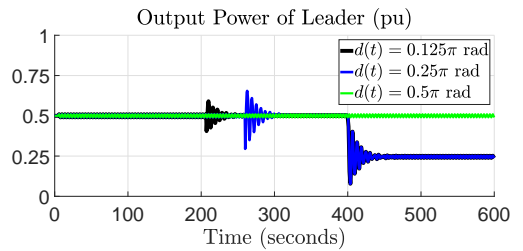}
			\caption{Power output of leader.\label{fig:power_dev_lead}}
		\end{figure}
		\begin{figure}[htbp!]
			\centering
			\includegraphics[width=0.75\columnwidth]{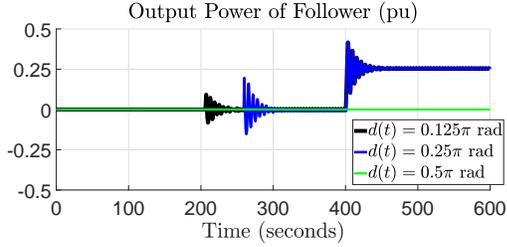}
			\caption{Power output of follower.\label{fig:power_dev_follow}}
		\end{figure}
		
		\subsection{High-order model testing}
		The simulation results we presented so far were developed using a reduced-order generator model that, in \cite{Ajala_NAPS2019}, was validated and shown to accurately mimic the behavior of a nineteenth-order model. In this section we provide additional results that show that our proposed synchronization method works when the same system is modeled using the nineteenth-order generator model.
		
		Figures \ref{fig:sync_errorMOD1} and \ref{fig:sync_errorMOD2} depict the observed synchronization error during the moment before the leader and follower systems are interconnected. Part (a) of the figure is for a case when the system is modeled using the second-order generator model, whereas Part (b) is for the case when the same system is modeled using the nineteenth-order generator model. For three disturbance values, i.e., $0$, $\frac{\pi}{32}$, and $\frac{\pi}{64}$ radians, the synchronization error is shown. For successful synchronization, the minimum threshold is $-0.134\pi$ rad, and the maximum threshold is $0.134\pi$ rad.
		Figures \ref{fig:phase_errorMOD1} and \ref{fig:phase_errorMOD2} show the observed phase synchronization error, wrapped around zero and $2\pi$, during the moment before the leader and follower systems are interconnected. Part (a) of the figure is for a case when the system is modeled using the second-order generator model, whereas Part (b) is for the case when the same system is modeled using the nineteenth-order generator model. The phase errors for three disturbance values, i.e., $0$ rad, $\frac{\pi}{32}$ rad, and $\frac{\pi}{64}$ rad, are shown. The maximum threshold for synchronization, $0.055\pi$ rad, is depicted using a red line.
		\begin{figure}[htbp!]
			\centering
			\includegraphics[width=\columnwidth]{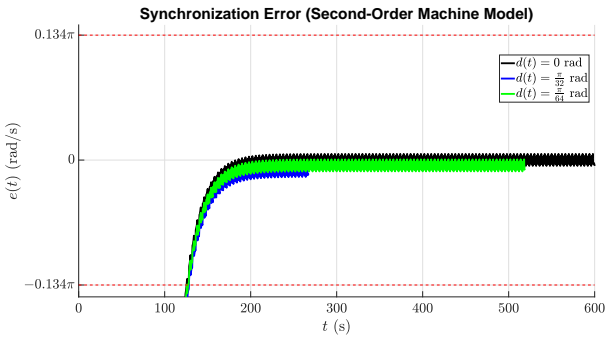}
			\caption{The observed synchronization error when a second-order generator model is used.\label{fig:sync_errorMOD1}}
		\end{figure}			
		\begin{figure}[htbp!]
			\centering
			\includegraphics[width=\columnwidth]{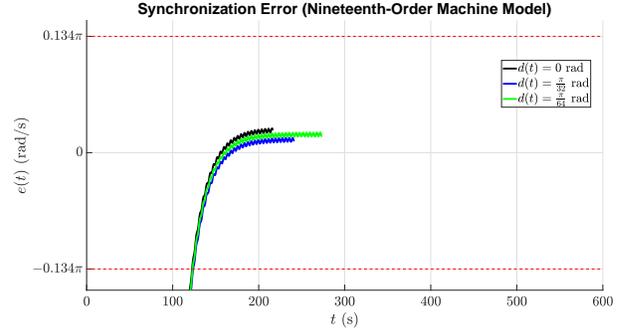}
			\caption{The observed synchronization error when a nineteenth-order generator model is used.\label{fig:sync_errorMOD2}}
		\end{figure}
		\begin{figure}[htbp!]
			\centering
			\includegraphics[width=\columnwidth]{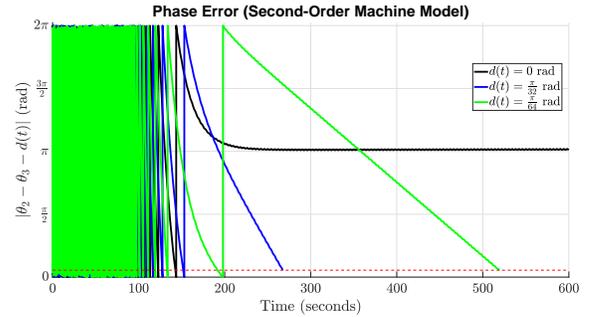}
			\caption{The observed phase error when a second-order generator model is used.\label{fig:phase_errorMOD1}}
		\end{figure}			
		\begin{figure}[htbp!]
			\centering
			\includegraphics[width=\columnwidth]{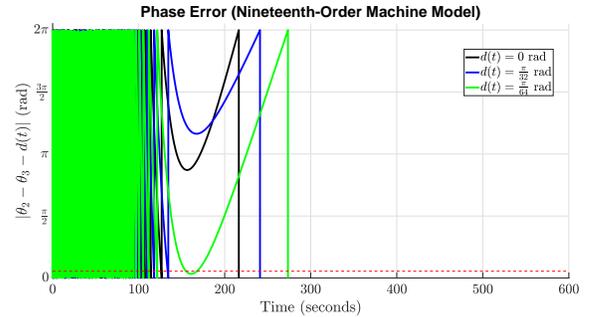}
			\caption{The observed phase error when a nineteenth-order generator model is used.\label{fig:phase_errorMOD2}}
		\end{figure}
		
		Figures \ref{fig:sync_errorMOD1} -- \ref{fig:phase_errorMOD2} depict that, for $d(t)=0$ rad, when the second-order model is employed, synchronization is unsuccessful, but when the nineteenth-order model is employed, synchronization is observed to be successful. This is due to the fact that, when the second-order model is employed, the observed phase error converges to a value close to $\pi$ rad, which is greater than the maximum threshold, but when the high-order generator model is employed, the phase error did not stay bounded when $d(t)=0$. The behavior of the high-order model results from the fact that the reduced-order model our synchronization method is developed from approximates the behavior of the high-order, and as a result introduces a model-reduction based disturbance to the control system of the high-order model based simulation.

		{\subsubsection*{Phase deviation error} According to synchronization requirements listed in \cite{thompson2012}, the magnitude of the phase deviation between the leader and follower must not exceed $0.055\pi$~rad. In other words,
			\begin{equation}\label{e-phase-dev1}
			\begin{split}
			\abs{\theta_2(t)-\theta_3(t)}\leq0.055\pi
			\end{split}
			\end{equation}
			must be enforced before the follower is connected to the leader. However, if measurements of $\theta_3(t)$ are corrupted by a disturbance $d(t)$, as described in Section \ref{s-system-description}, the detected magnitude of phase deviation becomes $\abs{\theta_2(t)-\theta_3(t)-d(t)}$, and the synchronization method would incorrectly enforce
			$$
			\abs{\theta_2(t)-\theta_3(t)-d(t)}\leq0.055\pi,$$ from where it follows that
			\begin{equation*}\label{e-phase-dev-d}
			\begin{split}
			\abs{\theta_2(t)-\theta_3(t)}\leq&\ 0.055\pi+\abs{d(t)},
			\end{split}
			\end{equation*} and for $\abs{d(t)}\leq0.25\pi$,
			\begin{equation}\label{e-phase-dev2}
			\begin{split}
			\abs{\theta_2(t)-\theta_3(t)}\leq&\ 0.305\pi.
			\end{split}
			\end{equation}
			The effect of a phase deviation error on the system frequency, the leader output power, and the follower output power is depicted in Figs. \ref{fig:bus_freq}, \ref{fig:power_dev_lead} and \ref{fig:power_dev_follow}. Although overcurrent protection devices are not modeled in the results, it is observed that, for a maximum disturbance of $d(t)=0.25\pi$, the transient of the post-synchronization system frequency slightly overshoots the permissible limits but quickly settles back within the limits.}


		\section{Phase-dependent damping}
		\label{s-phase-depend-damp}
		In this section we briefly consider the case when the leader model takes the form
		\begin{align}
		\dot \theta_1&=\omega_1,\\
		\dot \omega_1&=u_1-D_1(\theta_1)\omega_1+\xi_1(t)\label{e-leader-omega1-PD}
		\end{align}
		where $u_1$ is the control input as before and $\xi_1(t)$ is a signal which we assume for simplicity to be known. In contrast with the model~\eqref{eqn:powerangleD}, \eqref{eqn:gen_ang_speed1} considered earlier in the paper, here the damping $D_1(\cdot)$ is phase-dependent, which can arise, e.g., from modeling phase-dependent friction due to eccentricity of the generator rotor. We note that for the earlier model~\eqref{eqn:powerangleD}, \eqref{eqn:gen_ang_speed1}, $\xi_1(t)$ corresponds to the load $\ell(t)$, and having exact knowledge of the load makes the synchronization problem trivial. This is not the case, however, for the case of phase-dependent damping treated here, as we will see shortly. The goal of this section is to demonstrate the applicability of our synchronization method in a more general theoretical context which is only loosely related to the application scenario considered in the previous sections.
		
		The function
		$D_1(\cdot)$ is taken to be periodic with period $2\pi$ and to satisfy some mild assumptions, namely:
		\begin{enumerate}
			\item There exist numbers $\overline D_1>\underline D_1>0$ such that
			\begin{equation}\label{e-D1-under-over-line}
			\underline D_1\le D_1(r)\le  \overline D_1 \qquad\foral r\ge 0.
			\end{equation}
			
			\item There exists an $\eps>0$, sufficiently small, such that
			\begin{equation}\label{e-D1prime-eps}
			| D_1'(r)|\le \eps \qquad\foral r\ge 0.
			\end{equation}
		\end{enumerate}
		Later we will derive more specific constraints that $\underline D_1$, $\overline D_1$, and $\eps$ will have to satisfy. We also assume that $\dot\xi_1(t)$ is uniformly bounded and denote its upper bound by $M$:
		\begin{equation}\label{e-xi1dot-M}
		| \dot \xi_1(t)|\le M \qquad\foral t\ge 0.
		\end{equation}
		
		Next, we take the follower model to be of the form
		\begin{align*}
		\dot \theta_2&=\omega_2,\\
		\dot \omega_2&=u_2-D_2(\theta_2)\omega_2.
		\end{align*}
		The phase-dependent nature of $D_2$ is similar to that of $D_1$, but the exact form of the function $D_2(\cdot)$ is not important because it will be canceled by the control $u_2$.
		
		Here we are assuming, similarly to Section~\ref{s-system-description}, that measurements of the first state $\theta_1$ of the leader are corrupted by an additive disturbance $d(t)$ when being passed to the follower, while measurements of the second state $\omega_1$ are not available to the follower.
		
		\subsection{Control design and analysis}
		
		We define the control $u_1$ exactly as before by the equation~\eqref{e-Tm1-def}, where the dynamics of $\delta_1(t)$ are given by~\eqref{e-delta1-ode}.
		The closed-loop system (again, $(\omega_1,\delta_1)$-dynamics only) is now
		\begin{align*}
		\begin{pmatrix}\dot\omega_1\\
		\dot \delta_1\end{pmatrix}&=\begin{pmatrix}-D_1(\theta_1(t))\omega_1-k\delta_1+\xi_1(t)\\
		\omega_1-\omega_0\end{pmatrix}\\&
		=\begin{pmatrix}-D_1(\theta_1(t))&-k\\
		1 &0\end{pmatrix}
		\begin{pmatrix}\omega_1\\
		\delta_1\end{pmatrix}+\begin{pmatrix}\xi_1(t)\\
		-\omega_0\end{pmatrix}
		\end{align*}
		which we can view as a linear time-varying system driven by a time-varying perturbation that creates a time-varying equilibrium at
		\begin{equation}\label{e-equil-PD}
		\omega_1=\omega_0,\qquad \delta_1=\dfrac {\xi_1(t)-D_1(\theta_1(t))\omega_0}k=:\delta_0(t)
		\end{equation}
		(meaning that for each frozen time $t$, this is the equilibrium of the corresponding fixed affine system).
		Shifting the center of coordinates to this time-varying equilibrium by defining the variables
		$
		\bar\omega_1$ and $\bar\delta_1$ as in~\eqref{e-shift-center}, we obtain
		the dynamics
		\begin{align}
		\begin{pmatrix}\dot{\bar\omega}_1\\
		\dot{\bar\delta}_1\end{pmatrix}&=
		\begin{pmatrix}-D_1(\theta_1(t))&-k\\
		1 &0\end{pmatrix}
		\begin{pmatrix}\bar\omega_1\\
		\bar\delta_1\end{pmatrix}+\begin{pmatrix}0\\
		\nu(t)\end{pmatrix}\label{e-affine-TV}
		\end{align}
		where
		\begin{equation}\label{e-nu-PD}
		\nu(t):=\dfrac {D_1'(\theta_1(t))\omega_1(t)\omega_0-\dot\xi_1(t)}k.
		\end{equation}
		
		We now make the following observations:
		\begin{enumerate}
			\item The matrix
			\begin{equation}\label{e-A-of-t}
			{A}(t):=\begin{pmatrix}-D_1(\theta_1(t))&-k\\
			1 &0\end{pmatrix}
			\end{equation}
			is Hurwitz for each \emph{frozen} $t$.
			\item  Its time derivative
			\begin{equation}\label{e-dotA}
			\dot {{A}}(t)=\begin{pmatrix}-D_1'(\theta_1(t))\omega_1(t)&0\\
			0 &0\end{pmatrix}
			\end{equation}
			is small because $D_1'$ was assumed to be small, as long as $\omega_1$ is kept bounded under the action of the control $u_1$.
			\item  The perturbation signal $\nu(t)$ is bounded for the same reason and also because $\dot \xi_1(t)$ is assumed to be bounded.
		\end{enumerate}
		Applying results on stability of slowly time-varying linear systems (see, e.g., \cite{xiaobin-automatica} and the references therein), we now show that solutions of the closed-loop system are bounded and converge to a small neighborhood of the time-varying equilibrium~\eqref{e-equil-PD}; the size of this neighborhood is determined by the size of the perturbation $\nu(t)$. This relies on the following well-known result on stability of linear time-varying systems (see, e.g., \cite[Theorem 3.4.11]{ioannou-sun-adaptive-book}; see also~\cite{xiaobin-automatica} for some extensions).
		
		\begin{Lemma}\label{t-LTV} Consider the LTV
			system
			\begin{equation}\label{e-LTV}
			\dot x={A}(t)x
			\end{equation}
			and assume that:
			\begin{itemize}
				\item
				${A}(t)$ is Hurwitz for each fixed $t$, and
				there exist constants $c,\lambda>0$ such that for all $t$ and
				$ s $ we have\footnote{Here $\|\cdot\|$ stands for the induced matrix
					norm corresponding to the Euclidean norm.}
				\begin{equation}\label{e-commar}
				\big\|e^{{A}(t) s }\big\|\le ce^{-\lambda s }.
				\end{equation}
				
				\item $A(\cdot)$ is $C^1$ and uniformly bounded: there exists an $L>0$ such that $\|{A}(t)\|\le
				L$ for all $t$.

				\item $\|\dot{{A}}(t)\|\le\mu$ for all $t$, where $\mu>0$ is sufficiently small.
			\end{itemize}
			Then the system~\eqref{e-LTV} is exponentially
			stable.
		\end{Lemma}
		
		From the proof of the above result given in~\cite{ioannou-sun-adaptive-book}, an upper bound on $\mu$ that guarantees stability is obtained as
		\begin{equation}\label{e-mu-bound}
		\mu<\frac{\beta_1}{2\beta_2^3}
		\end{equation}
		where
		$$
		\beta_1:=\dfrac1{2L},\qquad \beta_2:=\dfrac{c^2}{2\lambda}.
		$$
		We now develop numerical expressions for these quantities.

		In our setting, the matrices ${A}(t)$ are given by~\eqref{e-A-of-t} and $D_1(\cdot)$ is assumed to satisfy the lower and upper bounds~\eqref{e-D1-under-over-line}. Proceeding analogously to Section~\ref{ss-param-eval}, we can show that we can take the common stability margin (i.e., exponential decay rate) $\lambda$ and the overshoot constant $c$ appearing in~\eqref{e-commar} to be
		$$
		\lambda:=\frac12{\underline D_1},\quad
		c=\sqrt{\dfrac{k+1+\sqrt{(k-1)^2+\overline D_1^2}}{k+1-\sqrt{(k-1)^2+\overline D_1^2}}}\,.
		$$

		Next, we need to find an $L$ satisfying the second hypothesis in Lemma~\ref{t-LTV}. This is straightforward: $\|{A}(t)\|$ is the largest singular value of ${A}(t)$, which is the square root of the largest eigenvalue of the matrix
		$$
		{A}^T(t){A}(t)=\begin{pmatrix}
		(D_1(\theta_1(t)))^2+1&D_1(\theta_1(t))k\\
		D_1(\theta_1(t))k &0
		\end{pmatrix}
		$$
		and this largest eigenvalue is
		\begin{align*}
		&\frac12\Big(D_1(\theta_1(t))^2+1\Big)\\&+
		\frac12\Big(\sqrt{((D_1(\theta_1(t)))^2+1)^2+4D_1(\theta_1(t))^2k^2}\Big)
		\\\le&
		\frac12\Big(\overline D_1^2+1+
		\sqrt{\big(\overline D_1^2+1\big)^2+4\overline D_1^2k^2}\Big).
		\end{align*}
		Choosing some specific value for $k$, we obtain a value for $L$ by taking the square root of the last quantity.

		Furthermore, exponential stability of the LTV system~\eqref{e-LTV} means that its state transition matrix $\Phi(\cdot,\cdot)$ satisfies
		\begin{equation}\label{e-Phi-GES}
		\|\Phi(t,s)\|\le \bar c e^{-\bar\lambda(t-s)}
		\end{equation}
		for some $\bar c,\bar\lambda>0$. The proof of Lemma~\ref{t-LTV} in~\cite{ioannou-sun-adaptive-book} yields the following estimates for the overshoot $\bar c$ and decay rate $\bar\lambda$:
		$$
		\bar c:=\sqrt{\frac{\beta_2}{\beta_1}},\qquad \bar\lambda:=\frac1{\beta_2}-\frac{2\beta_2^2}{\beta_1}\mu
		$$
		where $\bar\lambda>0$ in light of~\eqref{e-mu-bound}.
		
		The actual system~\eqref{e-affine-TV} is the LTV system~\eqref{e-LTV} driven by the perturbation~\eqref{e-nu-PD}.
		It is well known and easy to show that, as long as the exponential stability bound~\eqref{e-Phi-GES} is valid, ${\bar c}/{\bar\lambda}$ is the system's $\mathcal L_\infty$-induced gain, and for bounded perturbations satisfying $|\nu(t)|\le\bar\nu$ $\foral t$ for some $\bar\nu>0$, the solutions of~\eqref{e-affine-TV} satisfy
		\begin{equation}\label{e-LTV-final-bound}
		\left|\binom{\bar\omega_1(t)}
		{\bar \delta_1(t)}\right|\le \bar c e^{-\bar\lambda t}\left|\binom{\bar\omega_1(0)}
		{\bar \delta_1(0)}\right|+\frac{\bar c}{\bar\lambda}\bar\nu
		\qquad \foral t\ge 0.
		\end{equation}
		In particular, $c\bar\nu/{\bar\lambda}$ is the ultimate bound on the norm of the solution in steady state:
		$$
		\limsup_{t\to\infty}\left|\binom{\bar\omega_1(t)}
		{\bar \delta_1(t)}\right|\le \frac{c\bar\nu}{\bar\lambda}.
		$$
		
		Now, we can finish the analysis as follows. Given some range of initial conditions and the desired range in which we want the solution of our system~\eqref{e-affine-TV} to belong, we can determine sufficiently small upper bounds $M$ and $\eps$ on $\dot\xi_1(t)$ and on $D_1'(\cdot)$, respectively, such that the magnitude of $\nu(t)$ in~\eqref{e-nu-PD} (which depends on these two upper bounds as well as on the chosen range of $\omega_1$ around $\omega_0$ and the control gain $k$) is upper-bounded by a small enough $\bar\nu$ so that~\eqref{e-LTV-final-bound} guarantees that the solution indeed remains in the desired range. Recalling~\eqref{e-dotA} and decreasing the upper bound on $D_1'(\cdot)$ further if necessary, we can always ensure that the last hypothesis of Lemma~\ref{t-LTV} holds with $\mu$ satisfying~\eqref{e-mu-bound}.

		For the follower, we define the control $u_2$ as
		\begin{align*}
		u_2=&\ \Big(D_2(\theta_2(t))-D_1\big(\theta_1(t)+
		d(t)\big)\Big)\omega_2(t)\\&- k\big(\theta_1(t)+d(t)\big)+ k\omega_0t+\xi_1(t).
		\end{align*}
		Similarly to Section~\ref{ss-control-design}, the aim of this control is to  correct the difference between the damping functions $D_1$ and $D_2$ and try to match the other terms on the right-hand side of the closed-loop dynamics~\eqref{e-leader-omega1-PD} for the leader  (modulo the disturbance). We can then write the closed-loop dynamics of the follower as
		\begin{align*}
		\dot \theta_2&=\omega_2,\\
		\dot\omega_2&=-D_1\big(\theta_1+d(t)\big)\omega_2+u_1-kd(t)
		+\xi_1(t).
		\end{align*}

		\subsection{Synchronization analysis}

		With $e:=\omega_2-\omega_1$ we have
		\begin{align*}
		\dot e=\dot \omega_2-\dot\omega_1=&-D_1(\theta_1+d)\omega_2
		+D_1(\theta_1)\omega_1-kd\\
		=& -D_1(\theta_1+d)\omega_2+D_1(\theta_1+d)\omega_1\\&-D_1(\theta_1+d)\omega_1
		+D_1(\theta_1)\omega_1-kd
		\\=&-D_1(\theta_1+d)e\\&-\Big(D_1(\theta_1+d)-D_1(\theta_1)\Big)\omega_1-kd.
		\end{align*}
		With $V(e):=\frac 12 e^2$ we have
		\begin{align}
		\dot V=& -D_1(\theta_1+d)|e|^2-kde\nonumber\\&-\Big(D_1(\theta_1+d)-
		D_1(\theta_1)\Big)\omega_1 e
		\\\le& -\underline D_1|e|^2+|e|\phi(|d|)
		\label{e-Vdot-TV}
		\end{align}
		where
		$$
		\phi(r):=\max_{(\theta_1,\omega_1)\in\Omega,\,|d|\le r}
		\big|\big(D_1(\theta_1+d)-D_1(\theta_1)\big)\omega_1\big|+kr
		$$
		and $\Omega$ is a bounded set in which $\theta_1$ (mod $2\pi$) and $\omega_1$ evolve.
		Rewriting~\eqref{e-Vdot-TV} as
		$$
		\dot V\le-\underline D_1|e|\Big(|e|-\frac {\phi(|d|)}{\underline D_1}\Big)
		$$
		we obtain
		$$
		|e|>\frac {\phi(|d|)}{\underline D_1}\quad\Rightarrow\quad \dot V<0
		$$
		which gives ISS from $d$ to $e$ with ISS gain function $\phi(\cdot)/{\underline D_1}$.
		This implies, in particular, that
		$$
		\limsup_{t\to\infty}|e(t)|\le \frac1{\underline D_1} {\phi\Big(\limsup\limits_{t\to\infty}|d(t)|\Big)}.
		$$
		The fact that the ISS gain depends on a compact set in which the state of the leader system evolves makes the synchronization error dynamics \emph{quasi-ISS} with respect to $d$, in the sense of~\cite{qDES-obs}. This situation is more subtle than the one we had in Section~\ref{ss-sync}.

		\section{Concluding Remarks}
		\label{s-conclusion}
		In this paper, we proposed a method for synchronizing two electric power generators, which is robust against disturbances in the measurements on which the method relies. Analytical and numerical results were used to validate the proposed robust synchronization method.
		
		\begin{appendix}
			\section{Model formulation}
			\label{appendix}
			In this section, the generator model presented in Section \ref{s-system-description}, the so-called damped model, is derived from a high-order model, using singular perturbation analysis. In the derivations that follow, all state variables are expressed in a $qd0$ reference frame (see \cite{KrauseWasynczuk2013} for more details on Park's $qd0$ transformation). We note that all model parameters and variables are scaled, and normalized using the per-unit system \cite{sauer2006power}.
			\subsection{The high-order model}
			\label{sec:high-order}
			Let $\Phi_{q_2}(t)$ and $E_{d'}(t)$ denote the \textit{flux linkages} of two damper windings aligned with the quadrature axis ($q$-axis) of the synchronous machine, let $\Phi_{d_1}(t)$ and $E_{q'}(t)$ denote the \textit{flux linkages} of a damper winding and a field winding, respectively, aligned with the direct axis ($d$-axis) of the synchronous machine, and let $I_{q}$ and $I_{d}$ denote the $q$-axis and $d$-axis components of the \textit{stator output current}, respectively. Then, the damper winding dynamics can be described by:		
			\begin{align}
			\tau_{q'} \dot{E}_{d'} =& -E_{d'}+\left(X_q-X_{q'}\right)\left(I_q-\frac{X_{q'}-X_{q''}}{(X_{q'}-X_{k})^2}\left(\Phi_{q_2}\right.\right.\nonumber\\&\left.\left.+(X_{q'}-X_{k})I_q+E_{d'}\right)\right),\label{eqn:slowdamp}\\
			\tau_{q''} \dot{\Phi}_{q_2} =& -\Phi_{q_2} -\left(X_{q'}-X_{k}\right)I_q - E_{d'},\label{eqn:fastdamp1}\\
			\tau_{d''} \dot{\Phi}_{d_1} =& -\Phi_{d_1} - \left(X_{d'}-X_{k}\right)I_d + E_{q'}(t),\label{eqn:fastdamp2}
			\end{align}
			where $X_{k}$ denotes the machine leakage reactance, $X_q$ denotes the $q$-axis component of machine stator reactance, $X_{q'}$ and $X_{d'}$ denote machine transient reactances, $X_{q''}$ denotes the machine sub-transient reactance, and $\tau_{q''}$, $\tau_{d''}$ and $\tau_{q'}$ denote time constants of the resulting dynamical system (see \cite{Library2d} for more details).
			
			Let $\Phi_{q}(t)$ and $\Phi_{d}(t)$ denote the $q$-axis and $d$-axis components of \textit{flux linkages} for the stator windings, respectively. Let ${\omega}_1(t)$ denote the machine angular speed, in radians per second, and let ${\delta}_1(t)$ denote the power angle of the synchronous machine in radians. At the electrical network bus, let $V_3$ and $\delta_3$ denote the voltage magnitude, in per unit, and the voltage phase relative to a reference frame rotating at the nominal frequency, in radians, respectively. Let \begin{math}
			V_{q}\coloneqq V_3\cos({\delta}_1-{\delta}_3)\text{ and } V_{d}\coloneqq V_3\sin({\delta}_1-{\delta}_3),
			\end{math} so that {${V_3} = \sqrt{\left({V}_q\right)^2+\left({V}_d\right)^2}$}. Then, the stator winding dynamics are described by:
			\begin{gather}
			\begin{aligned}
			\dot{{\delta}_1}=&\ {\omega}_1(t)-\omega_0,\\
			\frac{1}{\omega_0}\dot{\Phi}_{q} =& - \frac{{\omega}_1(t)}{\omega_0}\Phi_{d} + V_{q}+R_{s}I_{q},\\
			\frac{1}{\omega_0}\dot{\Phi}_{d} =&\ \frac{{\omega}_1(t)}{\omega_0}\Phi_{q} + V_{d}+R_{s}I_{d},
			\end{aligned}\label{eqn:stator_network1}
			\end{gather}
			and
			\begin{gather}
			\begin{aligned}
			\Phi_{q} =&\ \frac{X_{q'}-X_{q''}}{X_{q'}-X_{k}}\Phi_{q_2}(t)- \frac{X_{q''}-X_{k}}{X_{q'}-X_{k}}E_{d'}(t) -X_{q''}I_{q},\\
			\Phi_{d} =&\ \frac{X_{d'}-X_{d''}}{X_{d'}-X_{k}}\Phi_{d_1}(t)+ \frac{X_{d''}-X_{k}}{X_{d'}-X_{k}}E_{q'}(t)-X_{d''}I_{d},
			\end{aligned}\label{eqn:stator_network2}
			\end{gather}
			where $X_{d''}$ denotes a machine sub-transient reactance, $R_{s}$ denotes the per-phase stator winding resistance, and $\omega_0$ denotes the nominal frequency, in radians per second.
			
			Let $E_{f}(t)$ denote the output voltage of the machine's excitation system, let $U_{f}(t)$ denote the exciter control input, let $\bar{U}_{f}(t)$ denote the rate feedback variable of the voltage regulator (see \cite{sauer2006power}, pp. 71--72 for details). Then, the dynamics of the machine's excitation system can be described as follows:
			\begin{equation}
			\begin{split}
			\tau_{d'} \dot{E}_{q'} =& -\left(X_d-X_{d'}\right)\left(I_d-\frac{X_{d'}-X_{d''}}{(X_{d'}-X_{k})^2}\left(\Phi_{d_1}\right.\right.\\&\left.\left.+(X_{d'}-X_{k})I_d-E_{q'}\right)\right) + E_f- E_{q'},\\
			\tau_f \dot{E}_f =& -K_fE_f + U_f,\\
			\tau_u \dot{U}_f =& -U_f+K_u\bar{U}_{f}-\frac{K_u\bar{K}_u}{\bar{\tau}_u}E_f + K_u\left({V}_r-{V_1}\right),\\
			\bar{\tau}_u \dot{\bar{U}}_{f} =& -\bar{U}_{f} + \frac{\bar{K}_u}{\bar{\tau}_u}E_f,
			\end{split}\label{eqn:excitation}
			\end{equation}
			where $V_r$ denotes the reference voltage magnitude, $X_d$ denotes the $d$-axis component of machine stator reactance, $X_f$ denotes the field winding reactance, $K_f$, $\bar{K}_u$, and $K_u$ denote known constants of the machine's excitation system, and $\tau_{d'}$, $\tau_f$, $\bar{\tau}_u$, and $\tau_u$ denote time constants of the resulting dynamical system (see \cite{Library2d} for more details).
			
			Let $T_m(t)$ denote the \textit{mechanical torque output} of the generator, and let $P_u(t)$ denote the \textit{fuel valve position} of the diesel engine, which acts as the prime mover. For the speed governor system, let $P_{a_2}$ denote the output of its actuator, with $\dot{P}_{a_1} =\ P_{a_2}$, and let $P_{b_2}$ denote the output of its electric control box, with $\dot{P}_{b_1} =\ P_{b_2}$. Then, the speed control system of the machine can be described by:
			\begin{gather}
			\begin{split}
			M\dot{{\omega}_1} =&\ T_{m} - \Phi_{d}(t)I_{q} + \Phi_{q}(t)I_{d} - \tilde{D}_0{\omega}_1,\\
			\tau_m\dot{T}_{m} =& -T_{m} + P_{u},\\
			\dot{P}_{u} =&\ P_{a_1} + \tau_{4}P_{a_2},\\
			\tau_{a_2}\dot{P}_{a_2} =& - \frac{1}{\tau_{5}+\tau_{6}}\left(P_{a_1} - \kappa\left(P_{b_1}+\tau_{3}P_{b_2}\right)\right) -P_{a_2},\\
			\tau_{2}\dot{P}_{b_2} =& -P_{b_2}\\- \frac{1}{\tau_{1}}&\left(P_{b_1}-\frac{1}{\bar{D}_{0}\omega_0}\left(\tilde{u} - P_{u}\right)+\frac{1}{\omega_0}\left({\omega}_1-\omega_0\right)\right),
			\end{split}\label{eqn:governor_engine}
			\end{gather}
			where $\tau_{2}$, $\tau_{3}$ , $\tau_{4}$, $\tau_{5}$ and $\tau_{6}$ denote time constants of the control system, $\tau_{a_2}=\frac{\tau_{5}\tau_{6}}{\tau_{5}+\tau_{6}}$, $\kappa$ denotes a controller gain for the actuator, $\tilde{u}$ denotes the power change setting of the machine, $M$ denotes the inertia of the machine, $\tilde{D}_0$ denotes the friction and windage damping coefficient of the machine, $\tau_m$ denotes the time constant of the engine, and $\bar{D}_{0}=\frac{1}{R_D\omega_0}$, with $R_D$ denoting the droop coefficient.
			
			\subsection{Time-Scale Properties of the High-Order Model}\label{sec:observation}
			The following observations are based on standard parameter values obtained from synchronous machine models in \cite{sauer2006power,Kundur1994,KrauseWasynczuk2013,Degov1MOD}, and an eigenvalue analysis of these models.
			\begin{itemize}
				\item[O1.] the dynamics of $\Phi_{q_2}$, $\Phi_{d_1}$, $E_{d'}$, $\Phi_{q}$, $\Phi_{d}$, $E_{q'}$, $E_{f}$, $U_{f}$, $\bar{U}_{f}$, $T_{m}$, $P_{u}$, $P_{a_2}$, $P_{b_2}$, $P_{a_1}$ and $P_{b_1}$, are much faster than those of ${\omega}_1$ and ${\delta}_1$.
				\item[O2.] for $\epsilon=0.1$, the parameters $R_{s}$, $\tau_{q''}$, $\tau_{q'}$, $\frac{1}{\omega_0}$, $\tau_f$, $\tau_u$, $\bar{\tau}_u$, $\tau_m$, $\tau_{a_2}$, $\tau_{2}$, $\tau_{1}$, $(\tau_{5}+\tau_{6})$, $\frac{\tau_{5}\tau_{6}}{(\tau_{5}+\tau_{6})}$, $\frac{1}{\kappa R_D}$ are $\bm{\mathcal{O}}\left(\epsilon\right)$, and for ${\omega}_1\approx\omega_0$, we have that $\frac{{\omega}_1(t)}{\omega_0}=1+
				\bm{\mathcal{O}}\left(\epsilon\right)$.\footnote{For a function $\epsilon\mapsto f(\epsilon)$ defined on some subset of the real numbers containing 0, we write
					\begin{math}
					f(\epsilon)=\bm{\mathcal{O}}\left(\epsilon^i\right)
					\end{math} if and only if there exists a positive real number $k$ such that
					\begin{math}
					\abs{f(\epsilon)}\leq k\abs{\epsilon^i} \text{ as }\epsilon\to0.
					\end{math}}
				\item[O3.] the dynamics of $\Phi_{q}$, $\Phi_{d}$ are much faster than those of $\Phi_{q_2}$, $\Phi_{d_1}$, $E_{d'}$ and $E_{q'}$.
				\item[O4.] the dynamics of $\Phi_{q_2}$ and $\Phi_{d_1}$ are much faster than those of $E_{d'}$ and $E_{q'}$.
			\end{itemize}
			We refer to $\Phi_{q_2}$, $\Phi_{d_1}$, $E_{d'}$, $\Phi_{q}$, $\Phi_{d}$, $E_{q'}$, $E_{f}$, $U_{f}$, $\bar{U}_{f}$, $T_{m}$, $P_{u}$, $P_{a_2}$, $P_{b_2}$, $P_{a_1}$ and $P_{b_1}$ as the fast states, and ${\omega}_1$ and ${\delta}_1$ as the slow states.

			\subsection{The Damped Model}\label{sec:DampedModel}
			The damped model is formulated by replacing differential equations for the fast states with algebraic counterparts, referred to as approximate manifolds. First-order approximate manifolds are formulated for the damper winding states, and zero-order approximate manifolds are developed for the other fast states \cite{kokotovicSingular}. By using a first-order approximation for the damper windings manifolds, the effects of damper windings on the generator response are captured by the resulting reduced model.
			
			Starting with the states observed to have the fastest dynamics, i.e., $\Phi_{q}(t)$ and $\Phi_{d}(t)$, we formulate the following zero-order approximations by setting $R_s=0$, $\frac{1}{\omega_0}=0$ and $\frac{{\omega}_1(t)}{\omega_0}=1$:
			\begin{equation}
			\begin{split}
			\Phi_{q}(t) \approx&\ \Phi_{q,0}(t) = - V_3\sin\left({\delta}_1(t)-{\delta}_3\right),\\
			\Phi_{d}(t) \approx&\ \Phi_{d,0}(t) = V_3\cos\left({\delta}_1(t)-{\delta}_3\right).
			\end{split}\label{eqn:PhiZeroMan}
			\end{equation}
			Next, for the subsequent fastest states, $\Phi_{q_2}(t)$ and $\Phi_{d_1}(t)$, which are damper winding states, we derive a first-order approximation of their manifolds. Manifolds for $\Phi_{q_2}(t)$ and $\Phi_{d_1}(t)$ can be expressed as power series in $\tau_{q''}$ and $\tau_{d''}$, respectively, to give:
			\begin{equation}
			\begin{split}
			\Phi_{q_2}(t) =&\  \Phi_{q_2,0}(t) + \tau_{q''}\Phi_{q_2,1}(t)+\cdots,\\
			\Phi_{d_1}(t) =&\  \Phi_{d_1,0}(t) + \tau_{d''}\Phi_{d_1,1}(t)+\cdots,
			\end{split}\label{eqn:approxMANx}
			\end{equation}
			from where it follows that first-order approximations are given by:
			\begin{equation}
			\begin{split}
			\Phi_{q_2}(t) \approx&\  \Phi_{q_2,0}(t) + \tau_{q''}\Phi_{q_2,1}(t),\\
			\Phi_{d_1}(t) \approx&\  \Phi_{d_1,0}(t) + \tau_{d''}\Phi_{d_1,1}(t).
			\end{split}\label{eqn:approxMAN}
			\end{equation}
			Expressions for $\Phi_{q_2,0}(t)$, $\Phi_{q_2,1}(t)$, $\Phi_{d_1,0}(t)$ and $\Phi_{d_1,1}(t)$ are derived by substituting \eqref{eqn:PhiZeroMan} into \eqref{eqn:stator_network2} to give $I_{q}  = \frac{V_3\sin\left({\delta}_1(t)-{\delta}_3\right)}{X_{q''}} + \frac{X_{q'}-X_{q''}}{\left(X_{q'}-X_{k}\right)X_{q''}}\Phi_{q_2}(t)- \frac{X_{q''}-X_{k}}{\left(X_{q'}-X_{k}\right)X_{q''}}E_{d'}(t)$ and $I_{d}  =  -\frac{V_3\cos\left({\delta}_1(t)-{\delta}_3\right)}{X_{d''}} + \frac{X_{d'}-X_{d''}}{\left(X_{d'}-X_{k}\right)X_{d''}}\Phi_{d_1}(t)+ \frac{X_{d''}-X_{k}}{\left(X_{d'}-X_{k}\right)X_{d''}}E_{q'}(t)$, substituting the resulting expressions, and \eqref{eqn:approxMAN}, into \eqref{eqn:fastdamp1}, \eqref{eqn:fastdamp2}, and equating the $\left(\tau_{q''}\right)^0$, $\left(\tau_{d''}\right)^0$, $\left(\tau_{q''}\right)^1$ and $\left(\tau_{d''}\right)^1$ terms to give: \begin{equation}
			\begin{split}
			\Phi_{q_2,0}(t) =&-\frac{X_{k}}{X_{q'}}E_{d'}(t)-\frac{X_{q'}-X_{k}}{X_{q'}}V_d,\\
			\Phi_{d_1,0}(t) =&\ \frac{X_{k}}{X_{d'}}E_{q'}(t)+\frac{X_{d'}-X_{k}}{X_{d'}}V_q,\label{eqn:manzeroCALC}
			\end{split}
			\end{equation}
			and
			\begin{equation}
			\begin{split}
			\Phi_{q_2,1}(t) =&-\frac{X_{q''}X_{k}}{\tau_{q'}X_{q'}^3}\left(X_{q}E_{d'}(t)-\left(X_{q}-X_{q'}\right)V_d\right)\\&+\dot{V}_d\frac{X_{q''}\left(X_{q'}-X_{k}\right)}{X_{q'}^2},\\
			\Phi_{d_1,1}(t) =&\ \frac{X_{d''}X_{k}}{\tau_{d'}X_{d'}^3}\left(X_{d}E_{q'}(t)-\left(X_{d}-X_{d'}\right)V_q\right)\\&-\frac{X_{d''}X_{k}}{\tau_{d'}X_{d'}^2}E_f(t)-\dot{V}_q\frac{X_{d''}\left(X_{d'}-X_{k}\right)}{X_{d'}^2},
			\end{split}\label{eqn:manoneCALC}
			\end{equation}
			where \begin{math}\dot{V}_{d}=V_3\cos({\delta}_1(t)-{\delta}_3)(\dot{{\delta}_1}(t)-\dot{\delta}_3)+\dot{V}_3\sin({\delta}_1(t)-{\delta}_3),\end{math} \begin{math}\dot{V}_{q}=\dot{V}_3\cos({\delta}_1(t)-{\delta}_3)-V_3\sin({\delta}_1(t)-{\delta}_3)(\dot{{\delta}_1}(t)-\dot{\delta}_3).\end{math}
			
			Next, for the damper winding state observed to have the slower dynamics, $E_{d'}(t)$, we derive a first-order approximation of its manifold. A manifold for $E_{d'}(t)$ can be expressed as a power series in $\tau_{q'}$ to give:
			\begin{equation}
			\begin{split}
			E_{d'}(t) =&\  E_{d',0}(t) + \tau_{q'}E_{d',1}(t)+\cdots,
			\end{split}\label{eqn:approxMANx2}
			\end{equation}
			from where it follows that a first-order approximation is given by:
			\begin{equation}
			\begin{split}
			E_{d'}(t) \approx&\  E_{d',0}(t) + \tau_{q'}E_{d',1}(t).
			\end{split}\label{eqn:approxMAN2}
			\end{equation}
			Expressions for $E_{d',0}(t)$ and $E_{d',1}(t)$ are derived by substituting \eqref{eqn:manzeroCALC} and \eqref{eqn:manoneCALC} into \eqref{eqn:approxMAN}, substituting the resulting expression, and \eqref{eqn:approxMANx2}, into \eqref{eqn:slowdamp} and equating the $\left(\tau_{q'}\right)^0$ and $\left(\tau_{q'}\right)^1$ terms to give:
			\begin{equation*}
			\begin{split}
			E_{d',0}(t) =&\ \frac{X_{q}-X_{q'}}{X_{q}}V_d-\frac{{N}_{q}}{{D}_{q}}\dot{V}_d,\\
			E_{d',1}(t) =&
			-\frac{{N}_{q'}}{\tilde{D}_{q}}\dot{V}_d + \bm{\mathcal{O}}\left(\tau_{q'}\right),
			\end{split}
			\end{equation*}
			with \begin{math}
			{N}_{q} = \tau_{q'}\tau_{q''}X_{q'}X_{k}(X_{q}-X_{q'})(X_{q'}-X_{q''})(X_{q'}-X_{k}),\; {D}_{q} = \tau_{q'}X_{q}X_{q'}^2(X_{q'}-X_{k})^2-\tau_{q''}X_{q}X_{k}^2(X_{q}-X_{q'})(X_{q'}-X_{q''}),\;
			{N}_{q'} = \tau_{q'}X_{q'}^3(X_{q}-X_{q'})(X_{q'}-X_{k})^2
			\end{math} and \begin{math}
			\tilde{D}_{q} = X_{q}{D}_{q}.
			\end{math}
			
			Finally, for other states observed to have fast dynamics, i.e., $E_{q'}$, $E_{f}$, $U_{f}$, $\bar{U}_{f}$, $T_{m}$, $P_{u}$, $P_{a_2}$, $P_{b_2}$, $P_{a_1}$, $P_{b_1}$, we derive zero-order approximations, while preserving the first-order approximations, by setting $\tau_{d'}$ and all $\bm{\mathcal{O}}\left(\epsilon\right)$ parameters identified in Section \ref{sec:observation}, except $\tau_{q''}$ and $\tau_{q'}$, to zero, to give:
			\begin{equation}
			\begin{split}
			E_{f,0}(t) =&\ \frac{K_u\left(V_r-V_1\right)}{K_f},\quad
			U_{f,0}(t) =\ K_fE_{f,0}(t),\\
			\bar{U}_{f,0}(t) =&\ \frac{\bar{K}_u}{\bar{\tau}_u}E_{f,0}(t),\quad
			P_{u,0}(t) =\ \tilde{u} - \bar{D}_{0}\left({\omega}_1(t)-\omega_0\right),\\
			T_{m,0}(t) =&\ P_{u,0}(t),\quad P_{a_1,0} = P_{a_2,0} = P_{b_1,0} = P_{b_2,0} = 0,\\
			E_{q',0}(t) =&\ \frac{X_{d'}}{X_{d}}E_{f,0}(t)+\frac{X_d-X_{d'}}{X_d}V_q-\frac{{N}_{d}}{{D}_{d}}\dot{V}_q,\\
			\end{split}\label{eqn:zeromanifold2}
			\end{equation}
			with \begin{math}
			{N}_{d} = \tau_{d'}\tau_{d''}X_{d'}{X_{k}}(X_{d}-X_{d'})(X_{d'}-X_{d''})(X_{d'}-X_{k}),\, {D}_{d} = \tau_{d'}X_{d}(X_{d'})^2(X_{d'}-X_{k})^2-\tau_{d''}X_{d}(X_{k})^2(X_{d}-X_{d'})(X_{d'}-X_{d''}).
			\end{math}	
			Substituting the first-order and zero-order approximate manifolds formulated in \eqref{eqn:approxMAN}, \eqref{eqn:approxMAN2} and \eqref{eqn:zeromanifold2} into \eqref{eqn:slowdamp}--\eqref{eqn:governor_engine}, and setting $\bm{\mathcal{O}}\left((\tau_{q'})^2\right)$ terms to zero, the damped model is given by:
			\begin{gather}
			\begin{split}
			\dot{{\delta}_1}=&\ {\omega}_1-\omega_0,\\
			M\dot{{\omega}_1} =&\ u_{1}-\frac{K_u\left(V_r-V_1\right)}{K_fX_{d}}V_d-\frac{\left(X_{d}-X_{q}\right)}{X_{q}X_{d}}V_qV_d\\&- C_{1}V_q\dot{V}_d +
			C_{2}V_d\dot{V}_q- {D}_{1}^{(0)}{\omega}_1,
			\end{split}\label{e-damped}
			\end{gather}
			where  $M$ denotes the scaled inertia constant of the machine, in seconds squared, $\omega_0$ denotes the nominal frequency, in radians per second, $K_u$ and $K_f$ denote excitation system constants, \begin{math}
			C_{1},\;C_{2}\end{math} are constants, with $C_{1}=C_{1''}+(C_{1'}+\tilde{C}_{1''})^2\tilde{C}_{1},\;
			C_{1''}=\frac{\tau_{q''}(X_{q'}-X_{q''})}{(X_{q'})^2},\;C_{1'}=\tau_{q'}X_{q'}(X_{q'}-X_{k}),\;\tilde{C}_{1''}=\frac{\tau_{q''}X_{q}X_{k}(X_{q'}-X_{q''})}{X_{q'}},\,\tilde{C}_{1}=\frac{(X_{q}-X_{q'})}{\tilde{D}_{q}},\;C_{2}=C_{2''}+(C_{2'}+\tilde{C}_{2''})\tilde{C}_{2''}\tilde{C}_{2},\;C_{2''}=\frac{\tau_{d''}(X_{d'}-X_{d''})}{(X_{d'})^2},\;C_{2'}=\tau_{d'}X_{d'}(X_{d'}-X_{k}),\;\tilde{C}_{2''}=\frac{\tau_{d''}X_{d}X_{k}(X_{d'}-X_{d''})}{X_{d'}},\;\tilde{C}_{2}=\frac{(X_{d}-X_{d'})}{\tilde{D}_{d}},\;\tilde{D}_{d} =X_{d}{D}_{d},
			$ and for $R_D$ denoting the frequency droop coefficient, $\tilde{u}$ denoting the power change setting of the generator, and $\tilde{D}_0$ denoting the friction and windage damping coefficient, {\begin{math}
				u_{1}=\tilde{u}+\frac{1}{R_D}\end{math} denotes the control input to the speed governor system, and \begin{math}{D}_{1}^{(0)}=\bar{D}_{0}+\tilde{D}_0
				\end{math}} are constants with $\bar{D}_{0}=\frac{1}{R_D\omega_0}$.
			
			Let $P$ denote the real power output of the synchronous generator, in per-unit. For the damped model, we have
			\begin{equation*}
			\begin{split}
			P=&\ \frac{K_u\left(V_r-V_1\right)}{K_fX_{d}}V_d+\frac{\left(X_{d}-X_{q}\right)}{X_{q}X_{d}}V_qV_d+ C_{1}V_q\dot{V}_d\\& - C_{2}V_d\dot{V}_q,
			\end{split}
			\end{equation*} and for the synchronous generator connected to a load with real power demand equal to $\ell(t)$, and reactive power demand equal to zero, it follows that $\ell(t)= P$.

			Assuming that the voltage support of the leader system is such that:
			\begin{inparaenum}[(i)]
				\item the voltage magnitude at the leader bus, i.e., $V_3$, is approximately constant at 1 per-unit, and
				\item the steady state voltage error of the leader, i.e., $V_{r}-V_1$, is approximately constant,
			\end{inparaenum} we choose $V_{r}-V_1=1$, and set $V_3=1$. It follows that, for ${\theta}_1 := {\delta}_1+\omega_0t$, and ${\theta}_3 := {\delta}_3+\omega_0t$, the damped model is described by:
			\begin{gather}
			\begin{split}
			\dot{\theta}_1=&\ {\omega}_1,\\
			M\dot{{\omega}_1} =&\ u_{1} - \ell(t) - {D}_{1}^{(0)}{\omega}_1,
			\end{split}\label{e-damped1}
			\end{gather}
			where
			\begin{math} \ell(t)=\ K_1\sin({\theta}_{1}-{\theta}_{3})+X_1\sin 2({\theta}_{1}-{\theta}_{3})+(C_{1}\cos^2({\theta}_{1}-{\theta}_{3}) +
			C_{2}\sin^2({\theta}_{1}-{\theta}_{3}))(\dot{{\theta}}_{1}-{\theta}_{3}),
			\end{math} with $K_1=\frac{K_u}{K_fX_{d}}$, and $X_1=\frac{(X_{d}-X_{q})}{2X_{q}X_{d}}$.
			Utilizing definitions in \eqref{e-Delta1-def}, \eqref{e-B1-def}, \eqref{e-D1-def}, and setting $M=1$, the model described by \eqref{eqn:powerangleD}, \eqref{eqn:gen_ang_speed1} and \eqref{e-load-new} can be reconstructed.

		\end{appendix}

		\bibliographystyle{IEEEtran}
		{\bibliography{shortstrings,TCNSrefs,lorenz,alejandro,mypapers}}
		
	\end{document}